\documentclass{article}
\usepackage{graphicx}
\usepackage{amsmath, amssymb}
\usepackage{natbib}
\usepackage{bm}

\bibpunct{(}{)}{;}{a}{,}{,}

\title{Scalar-on-function regression with measurement error in the functional regressors}

\author{Xiaochen Cai, 
Fenglin Xie, 
R.\ Todd Ogden \\
Department of Biostatistics, Columbia University, New York, New York, USA}

\date{}

\begin{document}

\maketitle

\begin{abstract}
We consider the problem of scalar-on-function regression. Most existing methods implicitly assume that the functional covariates are observed exactly, but in practice, they are often contaminated by measurement error. Our goal, therefore, is to deal with the problem of scalar-on-function regression when the regressor functions are observed with error. In this paper, we propose a simulation-extrapolation method to correct for the attenuation of estimated coefficient functions caused by the error. The method first estimates the error variance, establishes the relationship between a sequence of added error variance and the corresponding estimates of coefficient functions, and then extrapolates to the zero-error. We describe three methods to extrapolate the sequence of estimated coefficient functions. In a simulation study, we compare the performance of the simulation-extrapolation method with two pre-smoothing methods based on smoothing splines and functional principal component analysis. Next, we discuss the extension of the method in several directions, allowing for more complex noise covariance structures, multiple replications of functional predictors, generalized responses, and 2D and 3D functional predictors. Finally, we illustrate the methods by an application to diffusion tensor imaging data.
\end{abstract}

\section{Introduction}\label{intro}

We consider the problem of functional linear regression when the responses are scalar and the regressors are functions. Suppose that we observe \(\{X_{i}(t),Y_{i}\}_{i = 1,\dots,n}\), where \(X_{i}(t)\) are the ``true'' underlying functions, \(Y_{i}\) are responses. A scalar-on-function regression model \citep{Reiss2016} is given by
\[
Y_{i} = \beta_{0} + \int_{0}^{1} X_{i}(t)\beta(t)\,dt + \delta_{i}, \qquad i = 1,\dots,n, \tag{1}
\]
where \(\beta_{0}\) is the intercept, \(\beta(t)\) is the coefficient function evaluated at \(t\), and \(\delta_{i}\) are errors. It is usually assumed that \(\beta\) is smooth and the \(\delta_{i}\)s are iid normal with mean 0 and variance \(\sigma_{\delta}^{2}\). Most existing methods, for example, \citet{Cardot1999}, \citet{MarxEilers1999}, \citet{Cardot2003}, \citet{ReissOgden2007}, \citet{James2009}, implicitly assume that \(X_{i}(t)\) are observed exactly. However, in practice, they are often corrupted by some measurement error. Suppose that the observed curves are sampled over some discrete grid \(t_{1},\dots,t_{D}\), where \(D\) is the number of sampled points. A simple model to express measurement error in the functional regressors is given by
\[
\mathbf{Z}_{i} = \mathbf{X}_{i} + \boldsymbol{\epsilon}_{i}, \qquad i = 1,\dots,n, \tag{2}
\]
where \(\mathbf{Z}_{i} = (Z_{id})_{d=1,\dots,D}^{T}\) are the noisy observed curves, \(\mathbf{X}_{i} = (X_{i}(t_{d}))_{d=1,\dots,D}^{T}\) are the true curves, and \(\boldsymbol{\epsilon}_{i} = (\epsilon_{id})_{d=1,\dots,D}^{T}\) is a vector of measurement errors, assumed to be iid normal with mean \(\mathbf{0}\) and variance \(\sigma_{\epsilon}^{2}\mathbf{I}\). Note that this simple error assumption is made here for simplicity of development. In Section~\ref{extensions}, we will discuss several cases in which this assumption is unrealistic and develop methods to deal with those more general situations.

A reasonable pragmatic solution is to first pre-smooth the noisy observed curves \(\mathbf{Z}_{i}\) and then build the scalar-on-function model using the denoised curves in place of the true curves \(\mathbf{X}_{i}\). Pre-smoothing is commonly done by smoothing splines \citep{GreenSilverman1994, RamsaySilverman2005}. The observed curves could be smoothed together or individually, and the smoothing parameter may be chosen by generalized cross validation (GCV) \citep{CravenWahba1979} or restricted maximum likelihood (REML) \citep{Ruppert2003}. Alternatively, pre-smoothing may be done by functional principal component analysis (FPCA) \citep{StaniswalisLee1998, Yao2003, Yao2005, Goldsmith2011a}, kernel smoothing \citep{Nadaraya1964, Watson1964} or local polynomial regression \citep{FanGijbels1996}.

Pre-smoothing can be regarded as the first step in a two-step approach. Alternatively, one may consider estimating models (1) and (2) simultaneously. \citet{James2002} applied the expectation-maximization (EM) algorithm \citep{Dempster1977} to optimize the observed likelihood for models (1) and (2). The true curves \(\mathbf{X}_{i}\) can be represented in terms of spline bases and their spline coefficients can be treated as the unobserved variables in the EM algorithm. \citet{Cardot2007} generalized the total least squares (TLS) method \citep{VanHuffel1991}, which allows for simultaneous estimation of the coefficient and the true covariate when the covariate is univariate or multivariate in errors-in-variables linear model, to the context of functional linear regression. In their simulation study, the proposed method performed similarly to pre-smoothing with the Nadaraya-Watson kernel estimator. \citet{Goldsmith2011b} developed the variational Bayes method to jointly model the response and the observed curves via a penalized approach, which generalized Bayesian methods for errors-in-variables linear model to the functional data situation.

The approach that we develop here is quite different from the above approaches. It is based on the idea of the simulation-extrapolation (SIMEX) algorithm for classical linear regression, originally proposed by \citet{CookStefanski1994}, and thus it aims to correct for the bias in estimating the coefficient function induced by measurement error. In this paper, we will develop the SIMEX algorithm for scalar-on-function regression, which first estimates the error variance, establishes the relationship between a sequence of added error variance and the corresponding coefficient function estimates, and then extrapolates to zero-error. We introduce three methods to extrapolate the sequence of coefficient function estimates, which we term as linear, nonlinear, and local polynomial extrapolation. In a simulation study, we will compare the performance of the SIMEX algorithm with two pre-smoothing methods based on smoothing splines and FPCA. We will also introduce several other extensions. Recent methodological developments include a Bayesian semi-parametric approach using instrumental variables \citep{Zoh2024}, SIMEX-based extensions to generalized functional linear regression \citep{Luan2023a}, and functional quantile regression with error-prone covariates \citep{Chen2024quantile}.

The rest of this paper is organized as follows.  Section~\ref{methodology} proposes an adaptation of theSIMEX algorithm to scalar-on-function regression and introduces three methods to perform extrapolation. Section~\ref{asymptotic} explores the asymptotic behavior of the SIMEX estimator. Section~\ref{simulation}describes a simulation study to evaluate the performance of the proposed method. Section~\ref{extensions} discusses some extensions. Section~\ref{application} illustrates the methodology with an application to diffusion tensor imaging data, and Section~\ref{discussion} concludes with some discussion points.

\section{Methodology}\label{methodology}

\subsection{SIMEX in classical linear regression}\label{classical}

There is a vast amount of literature that deals with measurement error in regressors in classical linear regression (see \citet{Fuller1987} and \citet{Carroll2006}). Regression calibration and SIMEX are very generally applicable methods. They do not assume any parametric model for the distribution of the true covariate and can be regarded as the default approaches, due to their broad applicability \citep{Carroll2006}. The basic idea of regression calibration is to first obtain an estimate of the true covariate, by validation, instrument, or replicate data, and then replace the covariate with its estimate when building the linear regression model. In the functional context, the pre-smoothing approach is in the general spirit of calibration. In addition to regression calibration and SIMEX, the likelihood approach and the Bayesian approach are two general approaches, which each model the distribution of the true covariate parametrically. In the functional context, \citet{James2002} and \citet{Goldsmith2011b} can be regarded as generalizations of the likelihood approach and the Bayesian approach, respectively. 

In simple linear regression, measurement error in the covariate has an attenuation effect on the coefficient estimate \citep{Fuller1987}. Specifically, let \(Y_{i}\), \(X_{i}\), and \(Z_{i}\) be the response, the true covariate, and the observed covariate with measurement error, respectively. Under the simple linear regression model \(E(Y_{i}) = \beta_{0} + \beta X_{i}\) and the classical additive error model \(Z_{i} = X_{i} + \epsilon_{i}\),
\[
E(\hat{\beta}) = E\left(\frac{\sum_{i}(Z_{i} - \bar{Z})(Y_{i} - \bar{Y})}{\sum_{i}(Z_{i} - \bar{Z})^{2}}\right) = \beta \frac{\sigma_{X}^{2}}{\sigma_{X}^{2} + \sigma_{\epsilon}^{2}},
\]
where \(\sigma_{X}^{2}\) is the variance of the true covariate \(X_{i}\), and \(\sigma_{\epsilon}^{2}\) is measurement error variance. Thus, the coefficient estimate is biased towards zero and the bias increases as measurement error variance increases. Moving forward to multiple linear regression, however, the effect of measurement error on the coefficient estimates becomes complex \citep[see p.~63]{Carroll2006}. It could depend on the correlation between covariates and/or measurement errors.

To circumvent the complexity of analytically determining the effect of measurement error for multiple linear regression, SIMEX proposes to determine such an effect by Monte Carlo simulation \citep{CookStefanski1994}. Specifically, this is done by adding additional simulated measurement error to the covariate measured with error. The error variance of the resulting (noisier) covariates thus increases to \((1+\theta)\sigma_{\epsilon}^{2}\), where \(\theta\) controls the amount of added error. The choice of \(\theta\) has an effect on the corresponding estimator of \(\beta\), which we will denote by \(\hat{\beta}(\theta)\). The simulation is repeated a large number of times and the coefficient estimate \(\hat{\beta}(\theta)\) can be calculated by averaging over the simulations for a given \(\theta\). Thus, without explicitly deriving the expression for the effect of measurement error on the coefficient estimate \(E(\hat{\beta}(\theta))\), SIMEX estimates that as a function of \(\theta\) by simulation. The functional form of \(E(\hat{\beta}(\theta))\) is usually unknown, but can be approximated by fitting any of the following models:
\begin{itemize}
\item Linear model: \(E(\hat{\beta}(\theta)) = a + b\theta\);
\item Quadratic model: \(E(\hat{\beta}(\theta)) = a + b\theta + c\theta^{2}\);
\item ``Nonlinear'' model: \(E(\hat{\beta}(\theta)) = a + \frac{b}{\theta + c}\).
\end{itemize}
The ``nonlinear'' model is exact in some special cases, but the quadratic model remains attractive, because the ``nonlinear'' model requires the nonlinear least squares method to fit the model, which could be numerically unstable \citep[pp.~109-110]{Carroll2006}. Any of the above models can be then extrapolated to \(\theta = -1\), the case of zero-error, to get an estimate of \(\beta\) that would occur if there were no measurement errors in the predictor.

One major limitation of SIMEX is that it usually requires the error variance to be known. Given only observations \(\{Y_{i}, Z_{i}\}_{i=1,\dots,n}\) (without replicate data, etc.), we are generally not able to estimate \(\sigma_{\epsilon}^{2}\). Additional knowledge beyond the data needs to be known, but that is not a very common situation. Thus, SIMEX has some practical limitations for general data analysis. (We will show, however, that this limitation is readily overcome when the predictors are function-valued). Moreover, SIMEX was originally developed for the case when a single predictor variable is measured with error. To our knowledge, generalizing SIMEX for multiple covariates measured with error has not yet been accomplished. The main difficulties might be the limited utility of the three simple extrapolation methods, due to possibly high collinearity between covariates and/or possibly high correlation between measurement errors \citep[see p.~112]{Carroll2006}. In this work, we are interested in dealing with measurement error in functional regressors for scalar-on-function regression. The model resembles that for multiple linear regression. We generalize the idea of SIMEX to scalar-on-function regression and work to resolve the above limitations of SIMEX in the functional situation. A comprehensive review of SIMEX is provided by \citet{Sevilimedu2022}, and recent extensions include functional quantile regression \citep{Chen2024quantile} and high-dimensional error-prone data \citep{Chen2023}.

\subsection{SIMEX for scalar-on-function regression} 

Here we describe a functional SIMEX procedure for the simplest case: linear 1D scalar-on-function regression with the simplest variance assumptions. Several extensions will be described in Section~\ref{extensions}.

\subsubsection{General procedure}

In the scalar-on-function regression model (1), our interest focuses on estimating the coefficient function \(\beta\). SIMEX is a very general method that can be potentially applied together with most estimation methods for model (1), including \citet{Cardot2003}, \citet{ReissOgden2007}, \citet{Goldsmith2011a}, etc. Therefore, in the following presentation, we do not distinguish among the various estimation methods for \(\beta\). Let \(\hat{\boldsymbol{\beta}} = (\hat{\beta}(t_{d}))_{d=1,\dots,D}^{T}\) denote any estimator and let \(F\) denote the function that maps the range of the observed data to the parameter space of the coefficient function. The estimator obtained by disregarding the presence of measurement error, which we term the ``naive'' estimator, is
\[
\hat{\boldsymbol{\beta}}_{\mathrm{naive}} = F(\{\mathbf{Z}_{i}, Y_{i}\}_{i=1,\dots,n}).
\]

Next, we describe the simulation component of the SIMEX algorithm, which is intended to determine the relationship between measurement error and the coefficient function estimate. Consider adding additional measurement error to the observed curves \(\mathbf{Z}_{i}\):
\[
\mathbf{Z}_{\theta,i} = \mathbf{Z}_{i} + \sqrt{\theta}\,\boldsymbol{\epsilon}_{i},
\]
where \(\boldsymbol{\epsilon}_{i}\), ``pseudo random errors'' generated by Monte Carlo simulation, are iid normal with mean \(\mathbf{0}\) and variance \(\sigma_{\epsilon}^{2}\mathbf{I}\), and \(\theta\) controls the amount of added measurement error. Note that from equation (2), we have
\[
\operatorname{Var}(\mathbf{Z}_{i} \mid \mathbf{X}_{i}) = \sigma_{\epsilon}^{2}\mathbf{I}.
\]
Then the variance of the curves with added measurement error is
\[
\operatorname{Var}(\mathbf{Z}_{\theta,i} \mid \mathbf{X}_{i}) = (1+\theta)\operatorname{Var}(\mathbf{Z}_{i} \mid \mathbf{X}_{i}) = (1+\theta)\sigma_{\epsilon}^{2}\mathbf{I}, \tag{3}
\]
and thus the variance is inflated by a factor of \(1+\theta\).

To apply this algorithm, we would need to know (or at least have a good estimate of) the error variance \(\sigma_{\epsilon}^{2}\) in order to generate ``pseudo random errors''. As discussed in Section~\ref{classical}, this is viewed as a major limitation of the utility of the SIMEX algorithm in the context of classical linear regression. However, in the context of functional linear regression, this problem can be addressed, since the error variance, if unknown, may be estimated by well-established smoothing methods, e.g., smoothing splines \citep{GreenSilverman1994, RamsaySilverman2005}. The observed curves \(\mathbf{Z}_{i}\) can be smoothed together and the error variance estimate is given by
\[
\hat{\sigma}_{\epsilon}^{2} = \frac{\|\operatorname{vec}(\mathbf{Z}) - \operatorname{vec}(\hat{\mathbf{X}})\|^{2}}{nD - df(\lambda_{ss})},
\]
where \(\mathbf{Z}\) is a \(D\times n\) matrix whose \(i\)-th column is \(\mathbf{Z}_{i}\), \(\hat{\mathbf{X}}\) is a \(D\times n\) matrix whose \(i\)-th column is the smoothed curve \(\hat{\mathbf{X}}_{i}\), \(\lambda_{ss}\) is the smoothing parameter chosen by GCV, and \(df(\lambda_{ss}) = \operatorname{trace}(\mathbf{H})\) with \(\mathbf{H}\) being the projection matrix such that \(\operatorname{vec}(\hat{\mathbf{X}}) = \mathbf{H}\operatorname{vec}(\mathbf{Z})\).

After generating the curves with added ``pseudo random errors'', we estimate the corresponding coefficient function. To reduce the variation due to Monte Carlo simulation, we repeat the simulation and estimation step a large number of times and then take the average. We perform the simulation and estimation step for a sequence of values of \(\theta\) to examine the relationship between added measurement error and the coefficient function estimates. Then, for each \(\theta_{k} \in \{\theta_{1},\dots,\theta_{K}\}\) and for each simulation \(b \in \{1,\dots,B\}\), we have an estimate \(\hat{\boldsymbol{\beta}}_{b,k}\) given by
\[
\hat{\boldsymbol{\beta}}_{b,k} = F(\{\mathbf{Z}_{b,\theta_{k},i}, Y_{i}\}_{i=1,\dots,n}),
\]
where \(\mathbf{Z}_{b,\theta_{k},i}\) are the curves with added measurement error of amount \(\theta_{k}\), generated in the \(b\)-th simulation. Taking the average across estimates of \(B\) times, we obtain the estimate \(\hat{\boldsymbol{\beta}}_{k}\) corresponding to a given \(\theta_{k}\):
\[
\hat{\boldsymbol{\beta}}_{k} = \frac{1}{B}\sum_{b=1}^{B}\hat{\boldsymbol{\beta}}_{b,k}.
\]
Therefore, through Monte Carlo simulation, we can estimate the relationship between added measurement error with amount \(\theta_{k}\) and \(E(\hat{\boldsymbol{\beta}}_{k})\). This is the simulation component of the SIMEX algorithm.

Note that in equation (3), if we let \(\theta = -1\), then \(\operatorname{Var}(\mathbf{Z}_{\theta,i} \mid \mathbf{X}_{i}) = \mathbf{0}\). Since \(E(\mathbf{Z}_{\theta,i} \mid \mathbf{X}_{i}) = \mathbf{X}_{i}\), then \(E[(\mathbf{Z}_{\theta,i} - \mathbf{X}_{i})(\mathbf{Z}_{\theta,i} - \mathbf{X}_{i})^{T} \mid \mathbf{X}_{i}] = \mathbf{0}\). That is, \(\theta = -1\) corresponds to \(\mathbf{Z}_{\theta,i} = \mathbf{X}_{i}\), i.e., the case of no measurement error. So after determining the relationship between added measurement error with amount \(\theta_{k}\) and the corresponding \(\hat{\boldsymbol{\beta}}_{k}\), for \(\theta_{k} \in \{\theta_{1},\dots,\theta_{K}\}\), we would attempt to extrapolate the relationship back to \(\theta_{0} = -1\), the case of zero measurement error. This is the extrapolation component of the SIMEX algorithm, which can be understood hypothetically as removing the measurement error from the observed curves. Since the relationship between \(\theta_{k}\) and \(E(\hat{\boldsymbol{\beta}}_{k})\) is generally not known, extrapolation inherently results in an approximate estimation. (The Supplementary Materials provide a summary of the general procedure for application of SIMEX to the scalar-on-function setting.)

\begin{figure}[htb]
\centering
\includegraphics[width=\linewidth]{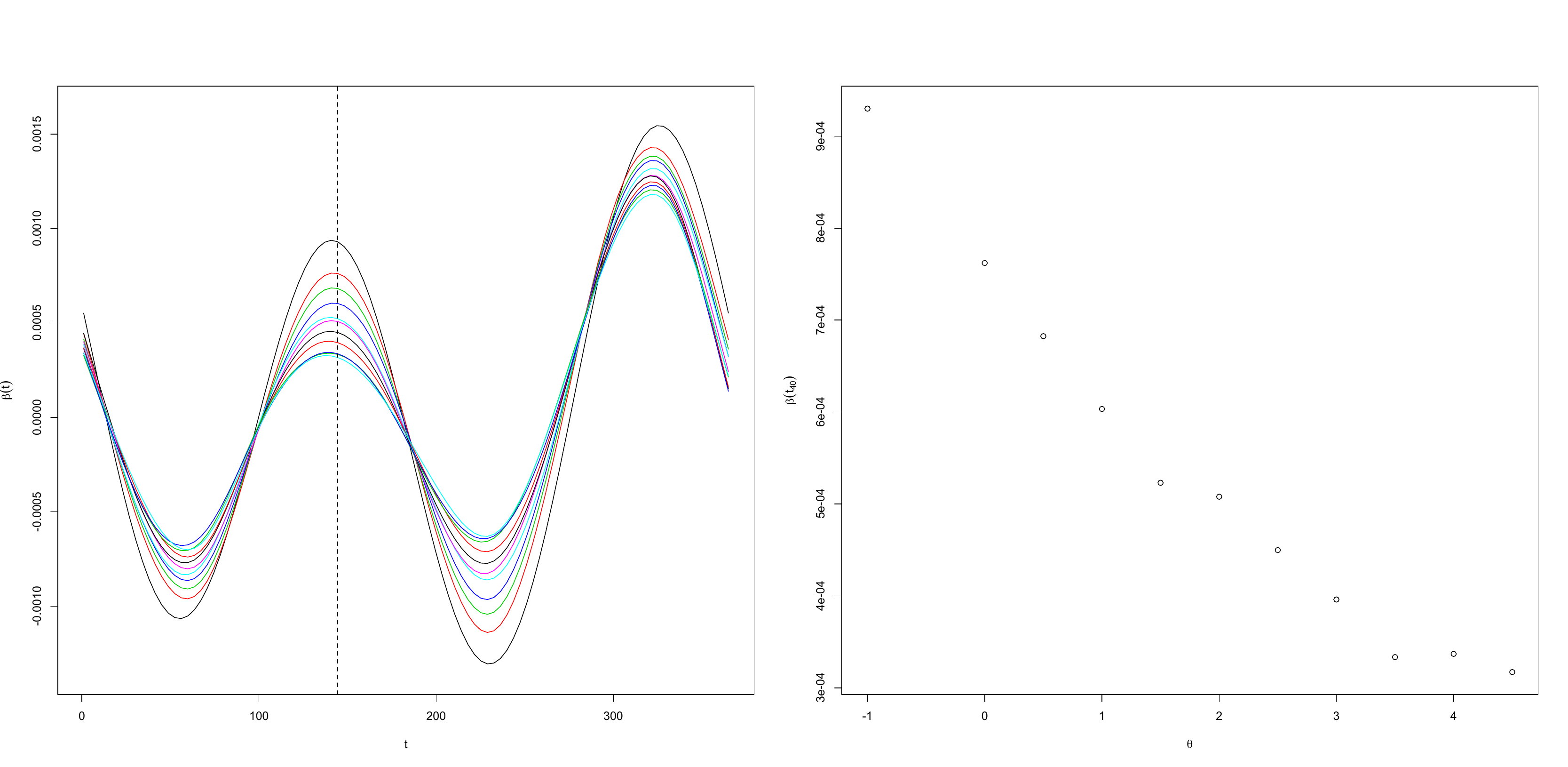}
\caption{An example of the relationship between added measurement error with amount $\theta_k$ and coefficient function estimates $\hat{\beta}_k$, for $\theta_k \in \{\theta_1,\dots,\theta_K\}$, in scalar-on-function regression. Left: the plot of $\hat{\beta}_k(t_d)$ vs.~$t_d$, for $\theta_k \in \{\theta_1,\dots,\theta_K\}$. The true coefficient function $\beta(t_d)$ is indicated by the solid black curve, the coefficient function estimates $\hat{\beta}_k(t_d)$ for sequentially increasing measurement error are indicated by lines of different colors, and a slice at $d=40$ is indicated by the dashed line. Right: the plot of $\hat{\beta}_k(t_{40})$ vs.~$\theta_k$. The true coefficient $\beta(t_{40})$ is plotted at $\theta_0=-1$.}
\label{fig:example1}
\end{figure}

Figure~\ref{fig:example1} gives an example of the relationship between added measurement error with amount \(\theta_{k}\) and coefficient function estimates \(\hat{\boldsymbol{\beta}}_{k}\), for \(\theta_{k} \in \{\theta_{1},\dots,\theta_{K}\}\). It is generated by one of the datasets from our simulation study, which we will describe in detail in Section~\ref{simulation}. It provides a nice graphical description of the attenuation effect in the coefficient function estimates caused by measurement error. Taking a slice of the coefficient function estimates at a given \(t\) and plotting those against the amount of measurement error \(\theta_{k}\), we can see that, for a given \(t\), the relationship is quite similar to what we expect to see in classic linear regression. Our aim in the next subsection is to explore the methods for extrapolating the relationship back to \(\theta_{0} = -1\), which corresponds to zero-error.

\subsubsection{Extrapolation methods}\label{extrapolation}

Here, we introduce three methods of extrapolation, which we term as linear, nonlinear, and local polynomial extrapolation.

\vskip 4mm
\noindent
\textbf{Linear extrapolation}
\vskip 2mm
\noindent
We consider fitting the following linear function-on-scalar regression model
\[
E(\hat{\beta}_{k}(t)) = b_{0}(t) + b_{1}(t)\theta_{k}, \qquad k = 1,\dots,K, \tag{4}
\]
where \(b_{0}(t)\) is the intercept function, and \(b_{1}(t)\) is the slope function. Let \(\hat{\boldsymbol{\beta}}(t) = (\hat{\beta}_{k}(t))_{k=1,\dots,K}^{T}\), \(\mathbf{b}(t) = (b_{m}(t))_{m=0,1}^{T}\) and \(\boldsymbol{\Theta} = [\mathbf{1}_{K}, (\theta_{k})_{k=1,\dots,K}^{T}]\), where \(\mathbf{1}_{K}\) is a length-\(K\) vector of 1's, and consider representing \(\hat{\boldsymbol{\beta}}(t)\) and \(\mathbf{b}(t)\) by a spline or some other basis expansion
\[
\hat{\boldsymbol{\beta}}(t) = \mathbf{C}\mathbf{B}(t), \qquad \mathbf{b}(t) = \mathbf{A}\mathbf{B}(t),
\]
where \(\mathbf{B}(t) = (B_{q}(t))_{q=1,\dots,Q}^{T}\) is a length-\(Q\) vector of basis functions evaluated at \(t\), \(\mathbf{C} = [C_{kq}]_{k=1,\dots,K; q=1,\dots,Q}\) is a \(K\times Q\) matrix of coefficients for \(\hat{\boldsymbol{\beta}}(t)\), and \(\mathbf{A} = [A_{mq}]_{m=0,1; q=1,\dots,Q}\) is a \(2\times Q\) matrix of coefficients for \(\mathbf{b}(t)\). Then model~(4) can be fitted by minimizing the following penalized least squares criterion \citep{RamsaySilverman2005, ReissHuang2010}:
\[
\int_{0}^{1} \| \mathbf{C}\mathbf{B}(t) - \boldsymbol{\Theta}\mathbf{A}\mathbf{B}(t) \|^{2} dt + \lambda \int_{0}^{1} [\mathbf{A}\mathbf{B}''(t)]^{T} \mathbf{A}\mathbf{B}''(t) dt, \tag{5}
\]
where \(\lambda\) is the tuning parameter, which may be chosen by minimizing the leave-one-curve-out CV criterion, defined as
\[
\frac{1}{K}\sum_{k=1}^{K} \int_{0}^{1} [\hat{\beta}_{k}(t) - \tilde{\beta}_{k}^{(-k)}(t)]^{2} dt,
\]
where \(\tilde{\beta}_{k}^{(-k)}\) is the predicted \(\hat{\beta}_{k}\), computed based on all curves except \(\hat{\beta}_{k}\).

\begin{figure}[htb]
\centering
\includegraphics[width=0.5\linewidth]{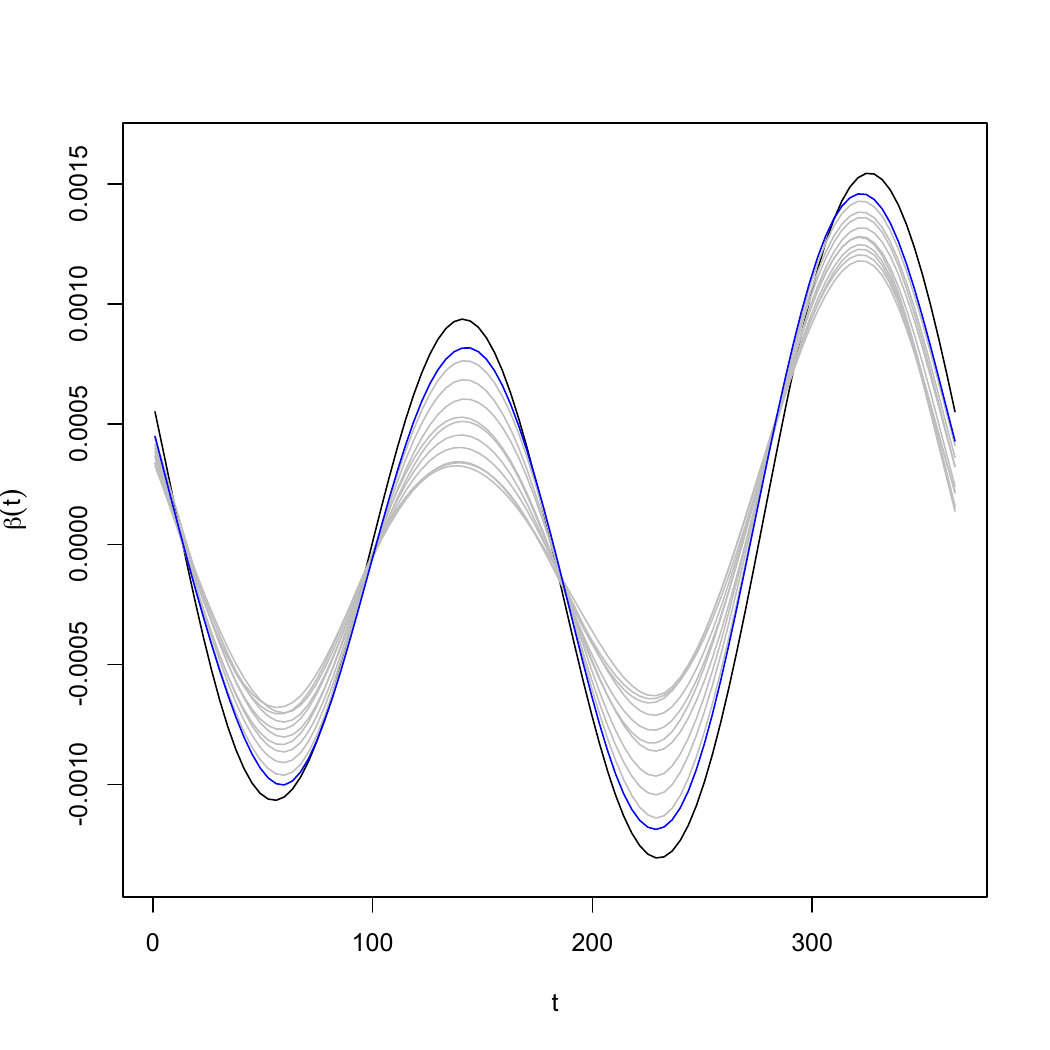} 
\caption{The coefficient function estimate by linear extrapolation. The true coefficient function $\beta(t)$ is indicated by the black curve. The coefficient function estimate by linear extrapolation $\hat{\beta}_{\mathrm{linear}}(t)$ is indicated by the blue curve. The coefficient function estimates $\hat{\beta}_k(t)$ for sequentially increasing measurement error are indicated by grey curves.}
\label{fig:linear}
\end{figure}

Note that in model (4), for a given \(t\), \(E(\hat{\beta}_{k}(t))\) has a linear relationship with \(\theta_{k}\), so we term this method of extrapolation as linear extrapolation. Following the example shown in Figure~\ref{fig:example1}, Figure~\ref{fig:linear} illustrates the coefficient function estimate obtained by performing linear extrapolation. The function-on-scalar regression model can also be extended to incorporate other functional forms of \(\theta_{k}\), such as a quadratic form.

\vskip 4mm
\noindent
\textbf{Nonlinear extrapolation}
\vskip 2mm
\noindent
We consider modeling \(\hat{\boldsymbol{\beta}}\) as a bivariate function of \(t\) and \(\theta\), that is
\[
E(\hat{\boldsymbol{\beta}}) = f(t,\theta), \tag{6}
\]
where \(f\) is a smooth bivariate function in \(\mathbb{R}^{2}\). The model can be fitted by extending the idea of one-dimensional smoothing splines to two dimensions, and the penalized least squares criterion is given by
\[
\|\hat{\boldsymbol{\beta}} - \mathbf{f}\|^{2} + J(f), \tag{7}
\]
where \(\hat{\boldsymbol{\beta}} = (\hat{\beta}_{dk})_{d=1,\dots,D; k=1,\dots,K}\), \(\mathbf{f} = (f(t_{d},\theta_{k}))_{d=1,\dots,D; k=1,\dots,K}\), and \(J\) is a penalty functional for functions in \(\mathbb{R}^{2}\). One common choice of \(J\) is
\[
J(f) = \lambda \int\int \left\{ \left(\frac{\partial^{2}f(t,\theta)}{\partial t^{2}}\right)^{2} + 2\left(\frac{\partial^{2}f(t,\theta)}{\partial t\partial\theta}\right)^{2} + \left(\frac{\partial^{2}f(t,\theta)}{\partial\theta^{2}}\right)^{2} \right\} dt\,d\theta, \tag{8}
\]
where \(\lambda\) is the tuning parameter. Minimizing (7) with \(J\) specified in (8) results in a thin plate spline smoother \citep{GreenSilverman1994}. When the degree of smoothness might be different between the two dimensions, which is likely to be the case for \(t\) and \(\theta\), a tensor product spline smoother with one roughness penalty per dimension might be more appropriate \citep{Wood2006b}. \(\mathbf{f}\) is represented by a tensor product spline basis expansion
\[
\mathbf{f} = (\mathbf{B}_{t} \otimes \mathbf{B}_{\theta})\mathbf{c},
\]
where \(\mathbf{B}_{t} = [B_{t,p}(t_{d})]_{d=1,\dots,D; p=1,\dots,P}\) is a \(D\times P\) matrix whose columns are \(P\) spline basis functions for dimension \(t\), \(\mathbf{B}_{\theta} = [B_{\theta,q}(\theta_{k})]_{k=1,\dots,K; q=1,\dots,Q}\) is a \(K\times Q\) matrix whose columns are \(Q\) spline basis functions for dimension \(\theta\), and \(\mathbf{c} = (c_{pq})_{p=1,\dots,P; q=1,\dots,Q}^{T}\) is a length-\(PQ\) vector of coefficients. Then minimizing criterion (7) with \(J\) given by
\[
J(f) = \int\int \left\{ \lambda_{t}\left(\frac{\partial^{2}f(t,\theta)}{\partial t^{2}}\right)^{2} + \lambda_{\theta}\left(\frac{\partial^{2}f(t,\theta)}{\partial\theta^{2}}\right)^{2} \right\} dt\,d\theta,
\]
where \(\lambda_{t}\) is the tuning parameter for dimension \(t\) and \(\lambda_{\theta}\) is the tuning parameter for dimension \(\theta\), results in a scale-invariant tensor product spline smoother \citep{Wood2006b}. The tuning parameters may be chosen by GCV or REML \citep{Wood2011}.

\begin{figure}[tbp]
\centering
\includegraphics[width=\linewidth]{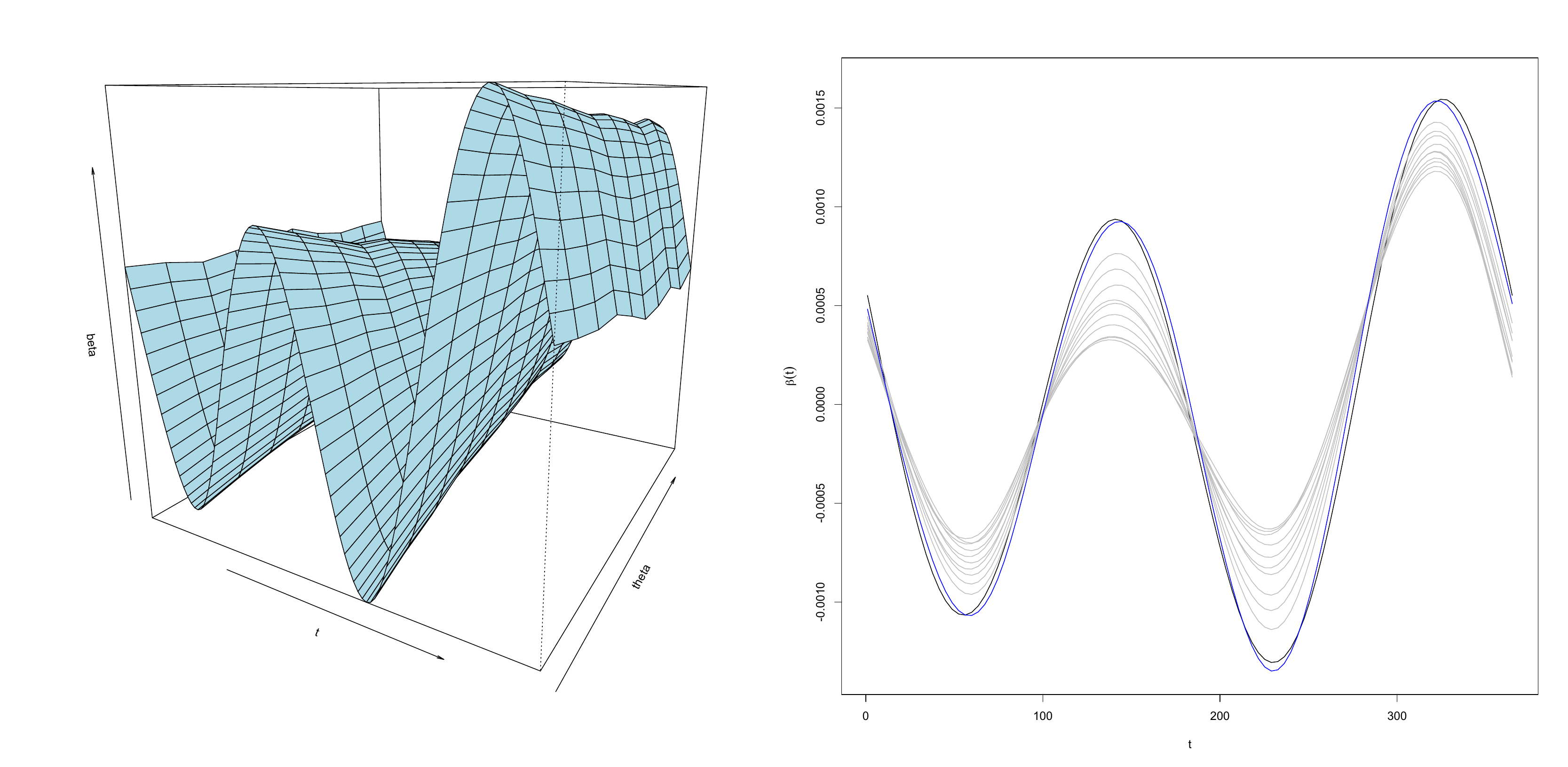}
\caption{The coefficient function estimate by nonlinear extrapolation based on a tensor product spline smoother. Left: the plot of $\hat{\beta}$ as a two-dimensional surface against $t$ and $\theta$. Right: the true coefficient function $\beta(t)$ is indicated by the black curve, the coefficient function estimate by nonlinear extrapolation $\hat{\beta}_{\mathrm{nonlin}}(t)$ is indicated by the blue curve, and the coefficient function estimates $\hat{\beta}_k(t)$ for sequentially increasing measurement error are indicated by grey curves.}
\label{fig:nonlinear}
\end{figure}

We term this method of extrapolation as nonlinear extrapolation. Continuing with the example in this section, the plot on the left of Figure~\ref{fig:nonlinear} illustrates \(\hat{\boldsymbol{\beta}}\) as a two-dimensional surface against \(t\) and \(\theta\). The plot on the right illustrates the coefficient function estimate obtained by performing nonlinear extrapolation based on a tensor product spline smoother.

\begin{figure}[htb]
\centering
\includegraphics[width=0.5\linewidth]{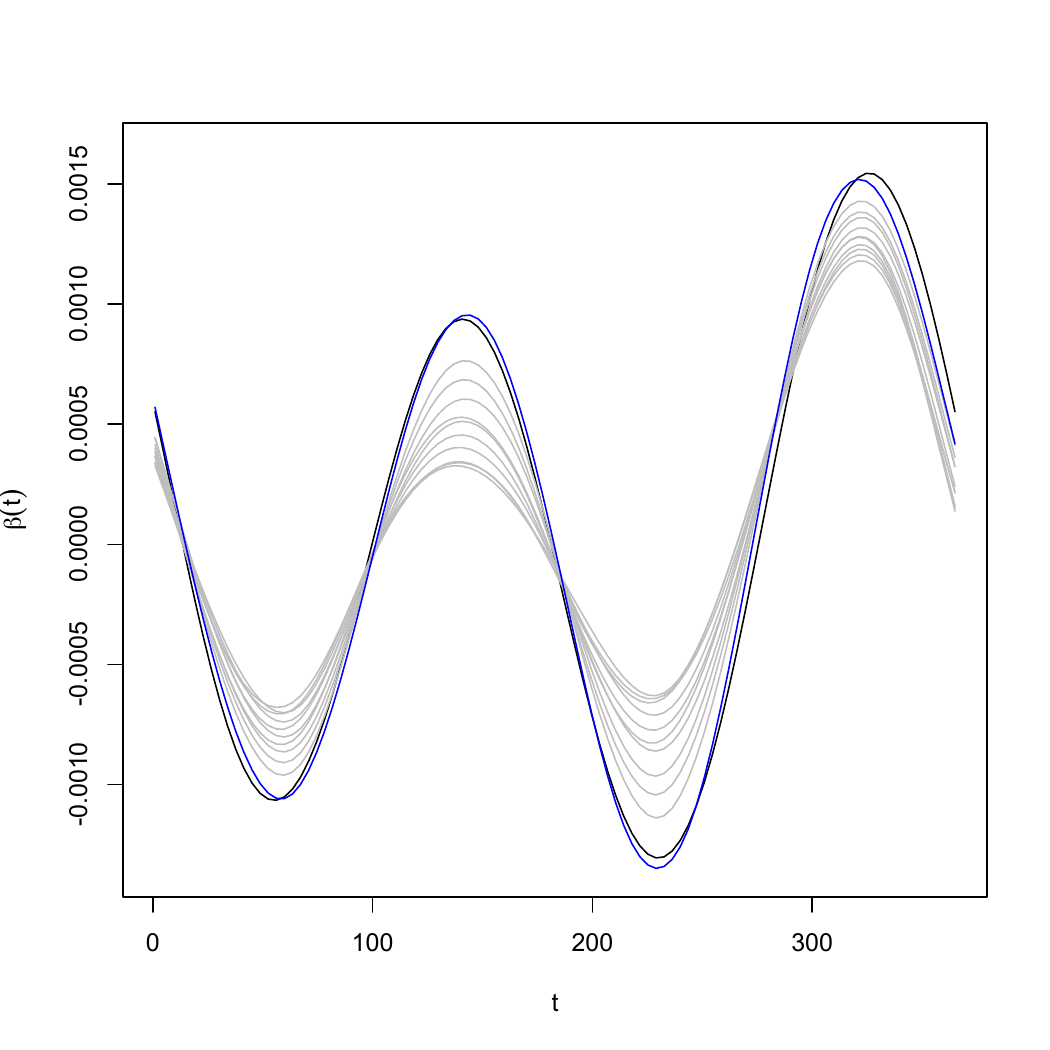} 
\caption{The coefficient function estimate by local polynomial extrapolation. The true coefficient function $\beta(t)$ is indicated by the black curve, the coefficient function estimate by local polynomial extrapolation $\hat{\beta}_{\mathrm{locpol}}(t)$ is indicated by the blue curve, and the coefficient function estimates $\hat{\beta}_k(t)$ for sequentially increasing measurement error are indicated by grey curves.}
\label{fig:local}
\end{figure}

\vskip 4mm
\noindent
\textbf{Local polynomial extrapolation}
\vskip 2mm
\noindent
Similar to nonlinear extrapolation, we consider modeling \(\hat{\boldsymbol{\beta}}\) as a smooth bivariate function of \(t\) and \(\theta\), that is,
\[
E(\hat{\boldsymbol{\beta}}) = f(t,\theta). \tag{9}
\]
We now introduce a hybrid approach of localized modeling and penalized basis expansion. In the \(\theta\) dimension, we consider fitting a localized model at the target point \(\theta_{0} = -1\), using only the curves that are close to the target point and weights determined using a kernel function. Meanwhile, we control the smoothness of the estimate as a function of \(t\), by incorporating a roughness penalty. Specifically, the model is fitted by minimizing the following penalized criterion
\[
\sum_{k=1}^{K} \int_{0}^{1} \left\{ \hat{\beta}_{k}(t) - \sum_{m=0}^{M} b_{m}(t)(\theta_{0} - \theta_{k})^{m} \right\}^{2} K_{h}(\theta_{0} - \theta_{k}) dt + \lambda \sum_{m=0}^{M} \int_{0}^{1} \{b_{m}''(t)\}^{2} dt, \tag{10}
\]
where \(M\) is the order of polynomial, \(K_{h}\) is a kernel function with bandwidth \(h\) and \(\lambda\) is the tuning parameter. Common choices of the kernel function include the Epanechnikov kernel \(K(x) = \frac{3}{4}(1-x^{2})_{+}\) and the Gaussian kernel \(K(x) = \frac{1}{\sqrt{2\pi}}\exp(-\frac{1}{2}x^{2})\). The bandwidth \(h\) is a tuning parameter controlling the smoothness in the \(\theta\) dimension. Let \(\hat{\boldsymbol{\beta}}(t) = (\hat{\beta}_{k}(t))_{k=1,\dots,K}^{T}\), \(\mathbf{b}(t) = (b_{m}(t))_{m=0,\dots,M}^{T}\), \(\boldsymbol{\Theta} = [(\theta_{0} - \theta_{k})^{m}]_{k=1,\dots,K; m=0,\dots,M}\) (a \(K\times(M+1)\) matrix), and \(\mathbf{K}_{h} = \operatorname{diag}(K_{h}(\theta_{0}-\theta_{1}), \dots, K_{h}(\theta_{0}-\theta_{K}))\). Consider representing \(\hat{\boldsymbol{\beta}}(t)\) and \(\mathbf{b}(t)\) by a spline or some other basis expansion
\[
\hat{\boldsymbol{\beta}}(t) = \mathbf{C}\mathbf{B}(t), \qquad \mathbf{b}(t) = \mathbf{A}\mathbf{B}(t),
\]
where \(\mathbf{B}(t) = (B_{q}(t))_{q=1,\dots,Q}^{T}\) is a length-\(Q\) vector of basis functions evaluated at \(t\), \(\mathbf{C} = [C_{kq}]_{k=1,\dots,K; q=1,\dots,Q}\) is a \(K\times Q\) matrix of coefficients for \(\hat{\boldsymbol{\beta}}(t)\), and \(\mathbf{A} = [A_{mq}]_{m=0,\dots,M; q=1,\dots,Q}\) is a \((M+1)\times Q\) matrix of coefficients for \(\mathbf{b}(t)\). Then the criterion (10) becomes
\[
\int_{0}^{1} [\mathbf{C}\mathbf{B}(t) - \boldsymbol{\Theta}\mathbf{A}\mathbf{B}(t)]^{T} \mathbf{K}_{h} [\mathbf{C}\mathbf{B}(t) - \boldsymbol{\Theta}\mathbf{A}\mathbf{B}(t)] dt + \lambda \int_{0}^{1} [\mathbf{A}\mathbf{B}''(t)]^{T} \mathbf{A}\mathbf{B}''(t) dt,
\]
which can be rewritten as
\[
\int_{0}^{1} \| \mathbf{C}^{*}\mathbf{B}(t) - \boldsymbol{\Theta}^{*}\mathbf{A}\mathbf{B}(t) \|^{2} dt + \lambda \int_{0}^{1} [\mathbf{A}\mathbf{B}''(t)]^{T} \mathbf{A}\mathbf{B}''(t) dt,
\]
where \(\mathbf{C}^{*} = \mathbf{K}_{h}^{1/2}\mathbf{C}\) and \(\boldsymbol{\Theta}^{*} = \mathbf{K}_{h}^{1/2}\boldsymbol{\Theta}\). Thus, the problem is converted to minimizing the penalized basis expansion criterion for a function-on-scalar regression model. The bandwidth \(h\) and the tuning parameter \(\lambda\) may be chosen by minimizing the leave-one-curve-out CV criterion
\[
\frac{1}{K}\sum_{k=1}^{K} \int_{0}^{1} [\hat{\beta}_{k}(t) - \tilde{\beta}_{k}^{(-k)}(t)]^{2} dt,
\]
We term this method of extrapolation as local polynomial extrapolation. Figure~\ref{fig:local} is an example of the coefficient function estimate obtained by performing local polynomial extrapolation.

\section{Asymptotic behavior}\label{asymptotic}

The asymptotics of the original SIMEX procedure were developed by \citet{Carroll1996}, who verified the asymptotic normality and computed the standard error estimates of the SIMEX estimator originally proposed by \citet{CookStefanski1994}. The original SIMEX procedure was intended for the classical linear regression situation involving only a single scalar-valued predictor measured with error. The extrapolation step was based on simple parametric regression.

In this section, we aim to explore the asymptotic behavior of SIMEX for scalar-on-function regression, considering only the case in which the extrapolation is based on linear extrapolation via function-on-scalar regression. Throughout this section, in the simulation step, the coefficient function estimation for a given amount of added measurement error \(\theta\) is based on the penalized spline estimator as in \citet{Cardot2003}, for which the penalized least squares criterion is given by
\[
\|\mathbf{Y} - \mathbf{Z}_{\theta}\mathbf{B}\gamma_{\theta}\|^{2} + \rho_{n}\gamma_{\theta}^{T}\mathbf{R}\gamma_{\theta}, \tag{11}
\]
where \(\mathbf{Y} = (Y_{1},\dots,Y_{n})^{T}\) is the response vector, \(\mathbf{Z}_{\theta} = [\mathbf{Z}_{\theta,1}, \dots, \mathbf{Z}_{\theta,n}]^{T}\) is a \(n\times D\) matrix of generated functional covariates with additional measurement error \(\theta\), \(\mathbf{B}\) is a fixed \(D\times p\) B-spline basis matrix, \(\gamma_{\theta}\) is a length-\(p\) vector of B-spline coefficients, \(\rho_{n}\) is the tuning parameter, and \(\mathbf{R}\) is the penalty matrix
\[
\mathbf{R} = \left[ \int B_{j}''(t) B_{j'}''(t) dt \right]_{j,j'=1,\dots,p}.
\]

For a sequence of \(\theta_{1},\dots,\theta_{K}\), having computed the B-spline coefficient estimates \(\hat{\gamma}_{S,\theta_{k}}\) (the subscript \(S\) indicates that the estimates are obtained through simulation), the extrapolation step is based on function-on-scalar regression with a least squares criterion given by
\[
\| (\mathbf{I}_{K}\otimes \mathbf{B})\hat{\boldsymbol{\gamma}}_{S,\Theta} - \operatorname{vec}(\mathbf{B}\mathbf{A}^{T}\boldsymbol{\Theta}^{T}) \|^{2}, \tag{12}
\]
where \(\hat{\boldsymbol{\gamma}}_{S,\Theta} = \operatorname{vec}(\hat{\gamma}_{S,\theta_{k}}, k=1,\dots,K)\), \(\boldsymbol{\Theta} = [\mathbf{1}_{K}, (\theta_{k})_{k=1,\dots,K}^{T}]\) is the design matrix and \(\mathbf{A} = [A_{mj}]_{m=0,1; j=1,\dots,p}\) is a \(2\times p\) B-spline coefficient matrix for the intercept and slope functions (note that the same fixed \(D\times p\) B-spline basis matrix \(\mathbf{B}\) is used as in the scalar-on-function regression model, for notational simplicity). Note that \(\theta_{0} = -1\) corresponds to the case of zero measurement error, so the true coefficient function \(\beta = \mathbf{B}\mathbf{A}^{T}\boldsymbol{\Theta}_{0}^{T}\), where \(\boldsymbol{\Theta}_{0} = (1, -1)\). To derive the asymptotic result for the SIMEX estimator, we make the following assumptions.

(A1) The observed functional predictors \(\mathbf{Z}_{i} = (Z_{id})_{d=1,\dots,D}^{T}\), for \(i=1,\dots,n\), are iid random vectors with \(E(Z_{id}) = 0\) and \(E(Z_{id}^{2}) < \infty\).

(A2) Measurement errors \(\boldsymbol{\epsilon}_{i} = (\epsilon_{id})_{d=1,\dots,D}^{T}\) are iid normal with \(E(\epsilon_{id}) = 0\) and known \(\sigma_{\epsilon}^{2} = E(\epsilon_{id}^{2}) < \infty\).

(A3) Given \(\theta \in [0,\infty)\), \(\mathbf{Y} - \mathbf{Z}_{\theta}\mathbf{B}\gamma_{\theta}\) is a vector of iid errors with mean 0 and finite variance, independent of \(\mathbf{Z}_{\theta}\).

(A4) The tuning parameter \(\rho_{n} = o_{p}(n^{1/2})\).

\vskip 3mm
\noindent
\textbf{Theorem 1} Suppose (A1)--(A4) hold. If \(\hat{\beta}_{\mathrm{SIMEX}}\) denotes a SIMEX estimate, then
\[
n^{1/2}(\hat{\beta}_{\mathrm{SIMEX}} - \beta) \xrightarrow{d} G,
\]
where \(G \sim N_{D}(\mathbf{0}, \boldsymbol{\Sigma})\) for some matrix \(\boldsymbol{\Sigma}\).

See Supplementary Materials for proof of Theorem~1.

\section{Simulation study}\label{simulation}

In this section, we perform a simulation study to compare the SIMEX algorithm with two pre-smoothing methods, including smoothing splines and FPCA, in terms of their performance in estimation and prediction.

\subsection{Setup}

\begin{figure}[htbp]
\centering
\includegraphics[width=0.5\linewidth]{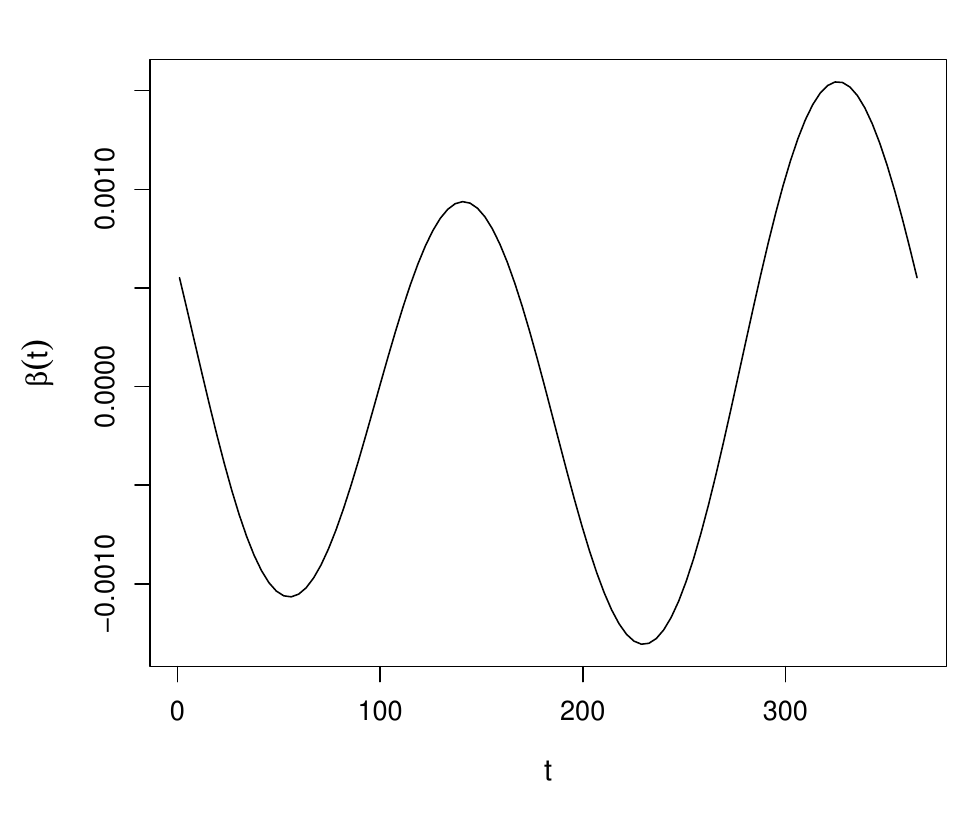}
\caption{The plot of the ``true'' coefficient function $\beta_k(t)$, which is generated by regressing the log of annual precipitation on the average daily temperature using Fourier basis functions.}
\label{fig:truebeta}
\end{figure}

We simulate \(N = 200\) samples of \(\{\mathbf{Z}_{i}, Y_{i}\}_{i=1,\dots,n}\) based on the following model
\[
Y_{i} = \mathbf{X}_{i}^{T}\boldsymbol{\beta} + \delta_{i}, \qquad i = 1,\dots,n,
\]
\[
\mathbf{Z}_{i} = \mathbf{X}_{i} + \boldsymbol{\epsilon}_{i}, \qquad i = 1,\dots,n,
\]
where \(\delta_{i}\) are iid normal with mean 0 and variance \(\sigma_{\delta}^{2}\) and \(\boldsymbol{\epsilon}_{i}\) are iid normal with mean \(\mathbf{0}\) and variance \(\sigma_{\epsilon}^{2}\mathbf{I}\). The true curves \(\mathbf{X}_{i}\) are generated based on the average daily temperature data at different locations over a year from the classic Canadian weather dataset \citep{RamsaySilverman2005, Ramsay2009}, available in the fda package in R \citep{Ramsay2014}. We first smooth the temperature data using smoothing splines and evaluate the smoothed data at \(D = 100\) evenly spaced time points \(t_{1},\dots,t_{D}\). Then we perform standard FPCA on the smoothed data and simulate \(n = 100\) curves as our true curves \(\mathbf{X}_{i}\) for each sample based on estimated eigenvalues and eigenfunctions \citep{Goldsmith2014}. The number of principal components is set to be the number of positive eigenvalues. Figure~\ref{fig:supp17} in the Supplementary Materials shows the average daily temperature data at different locations over time based on the original Canadian weather dataset and an example of the simulated curves \(\mathbf{X}_{i}\). The true coefficient function \(\beta\) is generated by regressing the log of annual precipitation on the average daily temperature using Fourier basis functions, as described by \citet{Ramsay2009} (see Figure~\ref{fig:truebeta}). 

The variance \(\sigma_{\delta}^{2}\) is set so as to ensure that \(R^{2} = \operatorname{Var}(\mathbf{X}_{i}^{T}\boldsymbol{\beta}) / (\operatorname{Var}(\mathbf{X}_{i}^{T}\boldsymbol{\beta}) + \sigma_{\delta}^{2}) = 0.95\). The standard deviation of the measurement error \(\sigma_{\epsilon}\) varies from 1.0 to 3.0 at intervals of 0.5. In addition, \(N_{\mathrm{test}} = 200\) test samples are generated in the same way as the training samples.

We are interested in comparing the following methods:
\begin{itemize}
\item \textbf{Oracle}: regressing \(Y_{i}\) on the true curves \(\mathbf{X}_{i}\) (note that the oracle method represents an ideal standard not achievable in practice, since the true curves \(\mathbf{X}_{i}\) are not known);
\item \textbf{Naive}: regressing \(Y_{i}\) on the observed curves \(\mathbf{Z}_{i}\);
\item \textbf{PS-Spline}: pre-smoothing by smoothing splines;
\item \textbf{PS-FPCA}: pre-smoothing by FPCA;
\item \textbf{SIMEX-Linear}: SIMEX with linear extrapolation, as described in Section~\ref{extrapolation};
\item \textbf{SIMEX-Nonlinear}: SIMEX with nonlinear extrapolation, as described in Section~3.2.2;
\item \textbf{SIMEX-LocPol}: SIMEX with local polynomial extrapolation, as described in Section~3.2.3.
\end{itemize}

The scalar-on-function regression is implemented as in \citet{Cardot2003}, using the code provided by the authors. The tuning parameter for the scalar-on-function regression \(\rho\) is set to range from \(1e^{-8}\) to \(9e^{-8}\) at intervals of \(1e^{-8}\), from \(1e^{-7}\) to \(9e^{-7}\) at intervals of \(1e^{-7}\), from \(1e^{-6}\) to \(9e^{-6}\) at intervals of \(1e^{-6}\), from \(1e^{-5}\) to \(9e^{-5}\) at intervals of \(1e^{-5}\) and from \(1e^{-4}\) to \(9e^{-4}\) at intervals of \(1e^{-4}\), chosen by GCV. The number of knots is 10.

For PS-Spline, the smoothing parameter ranges from \(1e^{0}\) to \(1e^{4}\) by a factor of 10, chosen by GCV. The number of knots is the same as the number of sampled time points.

For PS-FPCA, the bivariate smoother for smoothing the covariance matrix is a tensor product spline with tuning parameters chosen by REML. The number of knots at each dimension is 10. The number of principal components is set to be the number of positive eigenvalues (the estimate of the covariance matrix is not guaranteed to be nonnegative definite), which corresponds to 100\% of variance explained.

For SIMEX methods, the amount of added measurement error \(\theta_{k}\) ranges from 0 to 4.5 at intervals of 0.5 and the number of simulations for each \(\theta_{k}\) is \(B = 50\). The error variance \(\sigma_{\epsilon}^{2}\) is estimated by smoothing splines. The tuning parameter \(\rho\) used in estimation is kept the same for \(\theta_{1}\) through \(\theta_{K}\), so that \(\hat{\boldsymbol{\beta}}_{k}\) only varies through \(\theta_{k}\). For nonlinear extrapolation, the tensor product spline is constructed due to possibly different degree of smoothness between the two dimensions. The number of knots for each dimension is 10 and the tuning parameters are chosen by REML. For local polynomial extrapolation, we used the Epanechnikov kernel and the cubic polynomials. The bandwidth is set to be from 3 to 7 at intervals of 1, chosen by optimizing CV.

The methods above are compared in terms of their performance in estimation and prediction. The performance criterion for estimating \(\beta\) is standardized integrated squared error (ISE), given by
\[
\mathrm{ISE}(\beta) = \frac{(\boldsymbol{\beta} - \hat{\boldsymbol{\beta}})^{T}(\boldsymbol{\beta} - \hat{\boldsymbol{\beta}})}{\boldsymbol{\beta}^{T}\boldsymbol{\beta}}.
\]
Two performance criteria for predicting \(Y_{i}\) are in-sample prediction mean squared error (PMSEin)
\[
\mathrm{PMSEin}(Y) = \frac{1}{n}\sum_{i=1}^{n} (\mathbf{X}_{i}^{T}\boldsymbol{\beta} - \hat{Y}_{i})^{2},
\]
and out-of-sample prediction mean squared error (PMSEout)
\[
\mathrm{PMSEout}(Y) = \frac{1}{n}\sum_{i=1}^{n} (\mathbf{X}_{\mathrm{test},i}^{T}\boldsymbol{\beta} - \hat{Y}_{\mathrm{test},i})^{2},
\]
where \(\mathbf{X}_{\mathrm{test},i}\) and \(Y_{\mathrm{test},i}\) are based on the test samples not considered in the model fitting.

Table~\ref{tab:ise} gives the average \(\mathrm{ISE}(\beta)\) over 200 samples. SIMEX with nonlinear and local polynomial extrapolation generally outperforms SIMEX with linear extrapolation and the two pre-smoothing methods. Pre-smoothing with FPCA performs the worst in terms of estimation.

\begin{table}[tb]
\centering
\caption{Average $\mathrm{ISE}(\beta)$ over 200 samples.}
\label{tab:ise}
\begin{tabular}{l|ccccc}
\hline
 & $\sigma_{\epsilon}=1$ & $\sigma_{\epsilon}=1.5$ & $\sigma_{\epsilon}=2$ & $\sigma_{\epsilon}=2.5$ & $\sigma_{\epsilon}=3$ \\
\hline
Oracle & 0.1315 & 0.1443 & 0.1510 & 0.1296 & 0.1317 \\
Naive & 0.1367 & 0.1997 & 0.2789 & 0.3755 & 0.4363 \\
PS-Spline & 0.1370 & 0.2004 & 0.2743 & 0.3667 & 0.4367 \\
PS-FPCA & 0.1677 & 0.2528 & 0.3895 & 0.6535 & 0.8027 \\
SIMEX-Linear & 0.1278 & 0.1887 & 0.2729 & 0.3756 & 0.4489 \\
SIMEX-Nonlinear & 0.1632 & 0.1827 & 0.2212 & 0.3005 & 0.3584 \\
SIMEX-LocPol & 0.1823 & 0.1962 & 0.2201 & 0.2959 & 0.3408 \\
\hline
\end{tabular}
\end{table}

\begin{figure}[htbp]
\centering
\includegraphics[width=\linewidth]{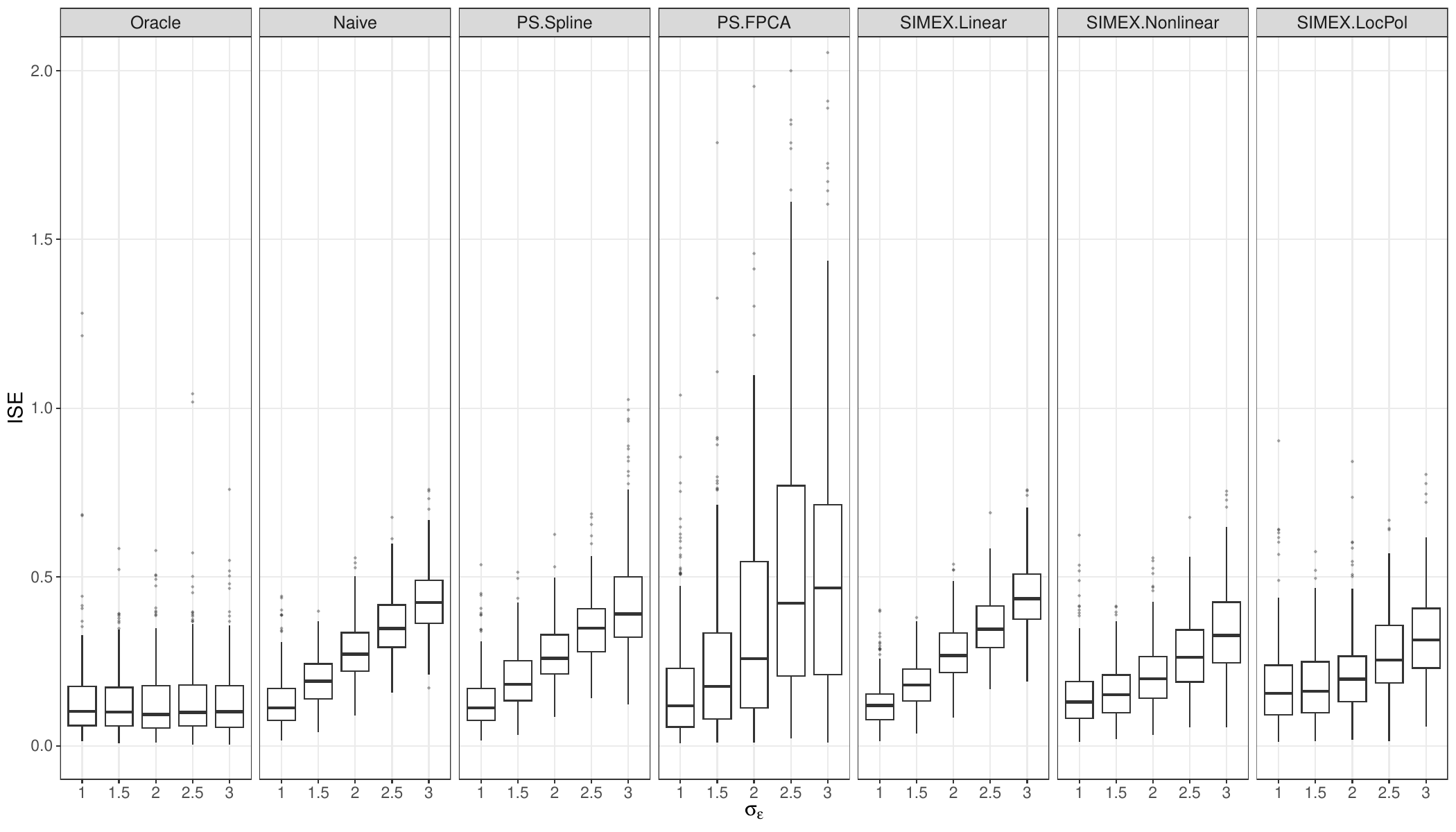}
\caption{Boxplots of $\mathrm{ISE}(\beta)$ over 200 samples. On the $x$-axis (from left to right) are Oracle, Naive, PS-Spline, PS-FPCA, SIMEX-Linear, SIMEX-Nonlinear, SIMEX-LocPol. Note that for PS-FPCA, the upper range of $\mathrm{ISE}(\beta)$ is truncated at 1.0.}
\label{fig:boxplot_ise}
\end{figure}

Figure~\ref{fig:boxplot_ise} shows the boxplots of \(\mathrm{ISE}(\beta)\) over 200 samples. Note that for PS-FPCA, the upper range of \(\mathrm{ISE}(\beta)\) is truncated at 1.0.

\begin{figure}[htbp]
\centering
\includegraphics[width=\linewidth]{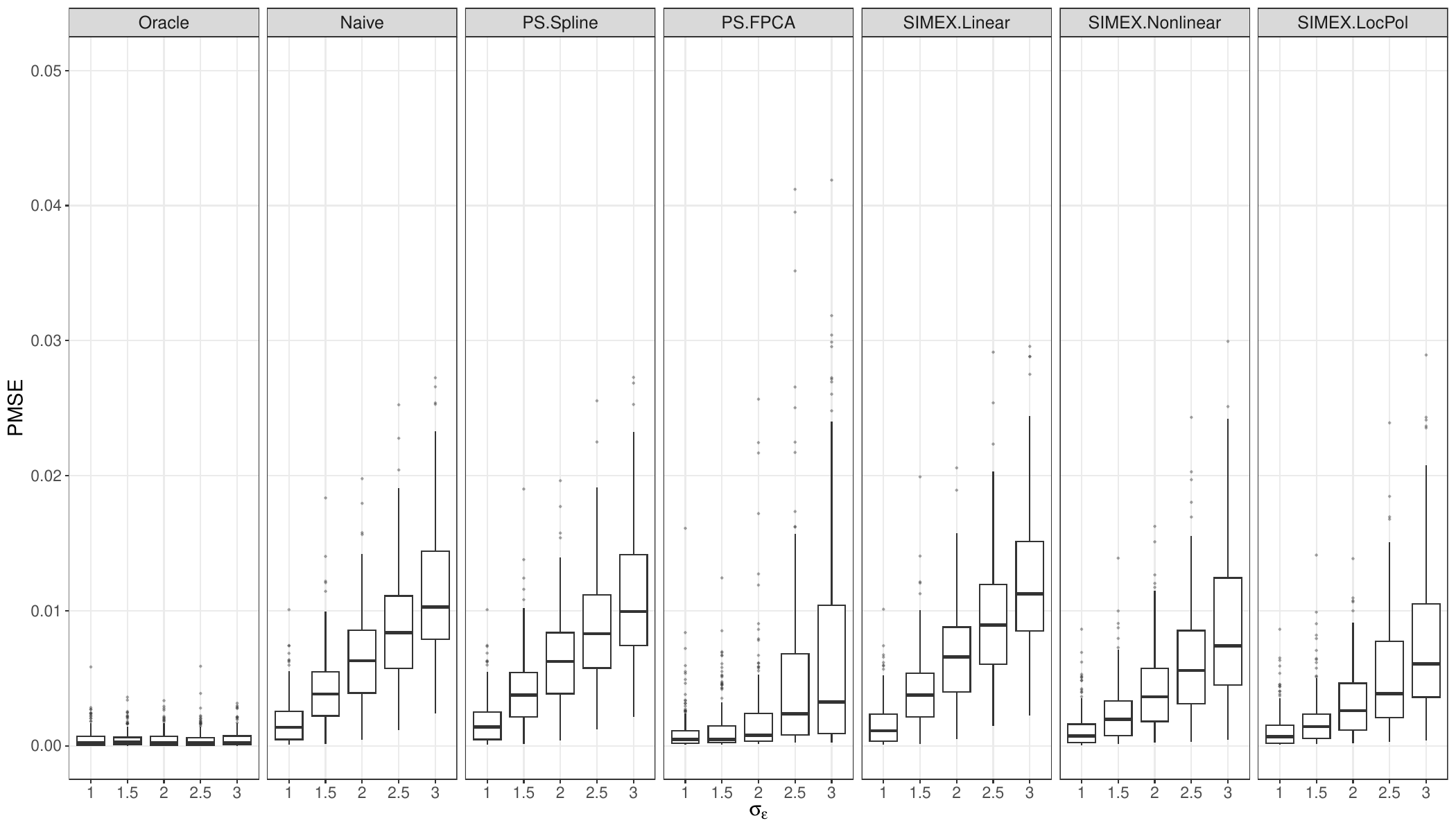}
\caption{Boxplots of $\mathrm{PMSEin}(Y)$ over 200 samples. On the $x$-axis (from left to right) are Oracle, Naive, PS-Spline, PS-FPCA, SIMEX-Linear, SIMEX-Nonlinear, SIMEX-LocPol. Note that for PS-FPCA, the upper range of $\mathrm{PMSEin}(Y)$ is truncated at 0.025.}
\label{fig:boxplot_pmsein}
\end{figure}

Table~\ref{tab:pmsein} gives the average \(\mathrm{PMSEin}(Y)\) over 200 samples. SIMEX with local polynomial extrapolation performs well, especially when the standard deviation of measurement error is large, i.e., \(\sigma_{\epsilon} = 2.5\) or 3.0. When the standard deviation is relatively small, pre-smoothing with FPCA and SIMEX with local polynomial extrapolation perform similarly well. Pre-smoothing with smoothing splines performs similarly to the naive method. Figure~\ref{fig:boxplot_pmsein} shows the box plots of \(\mathrm{PMSEin}(Y)\) for 200 samples. Although the range of \(\mathrm{PMSEin}(Y)\) for pre-smoothing with FPCA is quite broad (upper truncated at 0.025 in the boxplots) when the standard deviation of the measurement error is large, the medians are much lower than other methods (except for the oracle method). So, pre-smoothing with FPCA does not perform well in terms of estimation, but performs well in terms of prediction. This is a phenomenon that we will discuss in Section~8.

\begin{table}[htbp]
\centering
\caption{Average $\mathrm{PMSEin}(Y)$ over 200 samples.}
\label{tab:pmsein}
\begin{tabular}{l|ccccc}
\hline
 & $\sigma_{\epsilon}=1$ & $\sigma_{\epsilon}=1.5$ & $\sigma_{\epsilon}=2$ & $\sigma_{\epsilon}=2.5$ & $\sigma_{\epsilon}=3$ \\
\hline
Oracle & 0.0004 & 0.0005 & 0.0004 & 0.0005 & 0.0005 \\
Naive & 0.0015 & 0.0037 & 0.0061 & 0.0090 & 0.0110 \\
PS-Spline & 0.0015 & 0.0036 & 0.0060 & 0.0089 & 0.0106 \\
PS-FPCA & 0.0009 & 0.0020 & 0.0029 & 0.0064 & 0.0103 \\
SIMEX-Linear & 0.0014 & 0.0037 & 0.0063 & 0.0095 & 0.0119 \\
SIMEX-Nonlinear & 0.0010 & 0.0022 & 0.0036 & 0.0061 & 0.0082 \\
SIMEX-LocPol & 0.0009 & 0.0019 & 0.0030 & 0.0053 & 0.0070 \\
\hline
\end{tabular}
\end{table}

\begin{figure}[htbp]
\centering
\includegraphics[width=\linewidth]{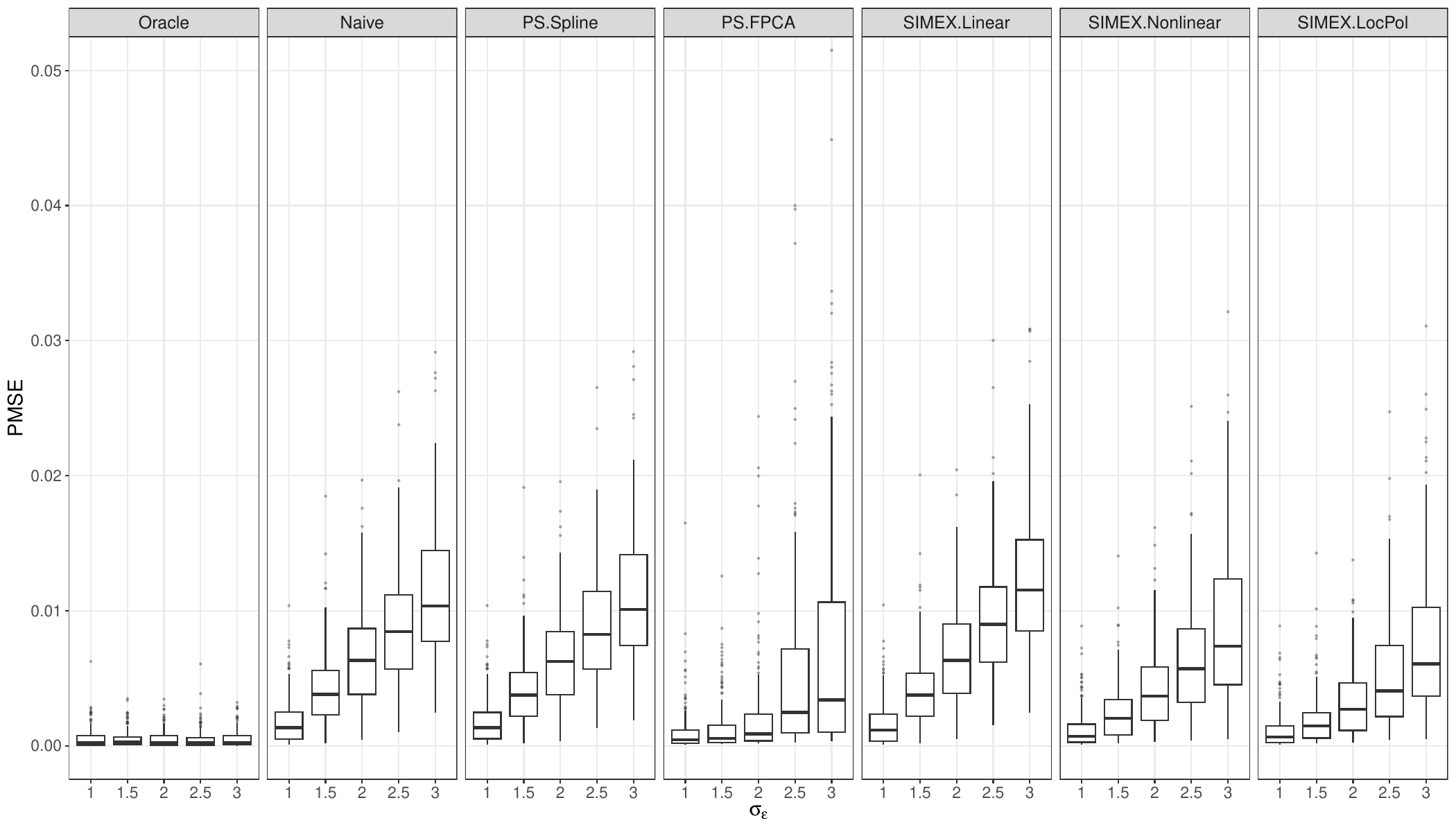}
\caption{Boxplots of $\mathrm{PMSEout}(Y)$ over 200 test samples. On the $x$-axis (from left to right) are Oracle, Naive, PS-Spline, PS-FPCA, SIMEX-Linear, SIMEX-Nonlinear, SIMEX-LocPol. Note that for PS-FPCA, the upper range of $\mathrm{PMSEout}(Y)$ is truncated at 0.025.}
\label{fig:boxplot_pmseout}
\end{figure}

Table~\ref{tab:pmseout} gives the average \(\mathrm{PMSEout}(Y)\) and Figure~\ref{fig:boxplot_pmseout} gives the boxplots of \(\mathrm{PMSEout}(Y)\) over 200 test samples. The pattern is very similar to what we observed for in-sample prediction, except that the average \(\mathrm{PMSEout}(Y)\) is slightly larger than the average \(\mathrm{PMSEin}(Y)\), as expected.

\begin{table}[htbp]
\centering
\caption{Average $\mathrm{PMSEout}(Y)$ over 200 test samples.}
\label{tab:pmseout}
\begin{tabular}{l|ccccc}
\hline
 & $\sigma_{\epsilon}=1$ & $\sigma_{\epsilon}=1.5$ & $\sigma_{\epsilon}=2$ & $\sigma_{\epsilon}=2.5$ & $\sigma_{\epsilon}=3$ \\
\hline
Oracle & 0.0004 & 0.0005 & 0.0004 & 0.0005 & 0.0005 \\
Naive & 0.0016 & 0.0039 & 0.0064 & 0.0094 & 0.0114 \\
PS-Spline & 0.0016 & 0.0039 & 0.0063 & 0.0093 & 0.0112 \\
PS-FPCA & 0.0009 & 0.0020 & 0.0030 & 0.0065 & 0.0103 \\
SIMEX-Linear & 0.0015 & 0.0039 & 0.0066 & 0.0099 & 0.0124 \\
SIMEX-Nonlinear & 0.0011 & 0.0023 & 0.0039 & 0.0065 & 0.0086 \\
SIMEX-LocPol & 0.0010 & 0.0021 & 0.0032 & 0.0057 & 0.0074 \\
\hline
\end{tabular}
\end{table}

In addition, the mean squared errors for estimating the standard deviation of measurement error over 200 samples are 0.0085, 0.0033, 0.0119, 0.0245, and 0.0123 when the true standard deviation of measurement error \(\sigma_{\epsilon} = 1.0, 1.5, 2.0, 2.5,\) and 3.0, respectively. Figure~\ref{fig:supp18} in the Supplementary Materials gives histograms of the estimated standard deviations of measurement error. The dashed line on each histogram indicates the true standard deviation of measurement error. We can see that the estimator \(\hat{\sigma}_{\epsilon}\) based on smoothing splines with smoothing parameter chosen by GCV tends to slightly underestimate the true standard deviation of the measurement error.

\section{Extensions}\label{extensions}

In this section, we will consider extending SIMEX in several directions. For the measurement error model (2), we made the assumption that errors are iid normal and the error variance is the same for all subjects. However, for most functional data, this assumption is too simple \citep[see p.~40]{RamsaySilverman2005}. The error variance could be subject-specific and errors could be autocorrelated and/or heteroscedastic. Here we incorporate more general assumptions on the covariance structure of measurement errors, which will be more realistic for functional data. We will consider the cases when the error variance is subject-specific and/or errors are autocorrelated. To ensure identifiability, the covariance structure is assumed to be specified by a parsimonious set of parameters. When the covariance structure is not parametrically specified, repeated measurements of functional data are required. So we will also discuss the case when there are repeated measurements made on the functional regressors. First, we will extend SIMEX to deal with the above cases. In addition, we will consider extending SIMEX to generalized functional linear regression, in order to incorporate response variables with a more general exponential family distribution. Finally, we will discuss extensions to 2D and 3D functional predictors.

\subsection{Subject-specific error variance}

When measurement errors within each observed curve can be assumed to be iid normal, but the error variance might vary by subject, we consider the following model for measurement error
\[
\mathbf{Z}_{i} = \mathbf{X}_{i} + \boldsymbol{\epsilon}_{i}, \qquad i = 1,\dots,n, \tag{13}
\]
where \(\boldsymbol{\epsilon}_{i}\) are measurement errors, assumed to be normal with mean \(\mathbf{0}\) and variance \(\sigma_{\epsilon,i}^{2}\mathbf{I}\). The subject-specific error variance \(\sigma_{\epsilon,i}^{2}\) can be estimated by applying smoothing methods separately on the observed curves for each subject. For example, by smoothing splines, the estimated error variance for subject \(i\) is given by \citep{Wahba1983}
\[
\hat{\sigma}_{\epsilon,i}^{2} = \frac{\|\mathbf{Z}_{i} - \hat{\mathbf{X}}_{i}\|^{2}}{D - df(\lambda_{ss})},
\]
where \(\hat{\mathbf{X}}_{i}\) is the smoothed curve by smoothing splines, \(D\) is the number of sampled points and \(\lambda_{ss}\) is the smoothing parameter chosen by GCV.

The rest of the SIMEX algorithm is the same as in Section~3, except that the variance of the pseudo errors to be added will depend on the subject-level variance:
\[
\mathbf{Z}_{\theta,i} = \mathbf{Z}_{i} + \sqrt{\theta}\,\boldsymbol{\epsilon}_{i}, \qquad i = 1,\dots,n,
\]
where \(\boldsymbol{\epsilon}_{i}\) are normal with mean \(\mathbf{0}\) and variance \(\sigma_{\epsilon,i}^{2}\mathbf{I}\). Then we have
\[
\operatorname{Var}(\mathbf{Z}_{\theta,i} \mid \mathbf{X}_{i}) = (1+\theta)\operatorname{Var}(\mathbf{Z}_{i} \mid \mathbf{X}_{i}) = (1+\theta)\sigma_{\epsilon,i}^{2}\mathbf{I}.
\]
Let \(\theta = -1\), then \(\operatorname{Var}(\mathbf{Z}_{\theta,i} \mid \mathbf{X}_{i}) = \mathbf{0}\). That is, \(\theta = -1\) corresponds the case of zero measurement error.

\subsection{Autocorrelated measurement error}

Most existing methods that can handle measurement error in functional regressors (for example, \citet{James2002}; \citet{Cardot2007}; \citet{Goldsmith2011a,Goldsmith2011b}) are based on an assumption that measurement errors are iid.  As mentioned earlier, such an assumption might be too simple for functional data. Measurement errors within a single observed curve often contain some amount of correlation in their covariance structure \citep[see p.~40]{RamsaySilverman2005}. Incorporating the presence of correlation, however, might pose difficulties in estimating the true function and the error function \citep{Opsomer2001}. On one hand, if the covariance structure is unstructured, i.e., not parametrically specified, the true function and the error function are not identifiable from each other. On the other hand, if the correlation is disregarded and the independent covariance structure is assumed, it might result in an underestimation of the smoothing parameter, when errors are actually positively correlated. The choice of the smoothing parameter is certainly critical for estimating the true function and the error function. Thus, to properly account for the correlation as well as to ensure identifiability, it is usually assumed that the covariance structure depends on a parsimonious set of parameters \citep{DiggleHutchinson1989, Kohn1992, Wang1998}.

Specifically, the measurement error model is given by
\[
\mathbf{Z}_{i} = \mathbf{X}_{i} + \boldsymbol{\epsilon}_{i}, \qquad i = 1,\dots,n, \tag{14}
\]
where \(\boldsymbol{\epsilon}_{i}\) are normal with mean \(\mathbf{0}\) and variance \(\sigma_{\epsilon,i}^{2}\mathbf{V}_{i}\). Here, we allow \(\sigma_{\epsilon,i}^{2}\) and \(\mathbf{V}_{i}\) to vary by subject. The correlation structure \(\mathbf{V}_{i}\) is dependent on a set of parameters, e.g., first-order autoregressive, compound symmetric, exponential correlation structure, etc. Suppose that we represent the true curve \(\mathbf{X}_{i}\) by a set of B-spline basis functions. Then model (14) can be fitted by minimizing the following generalized least squares criterion with a roughness penalty \citep[see p.~87]{RamsaySilverman2005}:
\[
(\mathbf{Z}_{i} - \mathbf{B}\gamma_{i})^{T}\mathbf{V}_{i}^{-1}(\mathbf{Z}_{i} - \mathbf{B}\gamma_{i}) + \lambda_{i}\gamma_{i}^{T}\mathbf{R}\gamma_{i}, \tag{15}
\]
where \(\mathbf{B} = [B_{k}(t_{d})]_{d=1,\dots,D; k=1,\dots,K}\) is a \(D\times K\) matrix whose columns are vectors of B-spline basis functions \(B_{k}(t)\) evaluated at \(t_{1},\dots,t_{D}\), \(\gamma_{i} = (\gamma_{ik})_{k=1,\dots,K}^{T}\) is a length-\(K\) vector of coefficients, \(\lambda_{i}\) is the smoothing parameter and \(\mathbf{R} = [\int B_{k}''(t)B_{k'}''(t)dt]_{k,k'=1,\dots,K}\) is a \(K\times K\) penalty matrix.

Criterion (15) can be related to that for a linear mixed effects model, so that \(\sigma_{\epsilon,i}^{2}\) and \(\mathbf{V}_{i}\) can be estimated as variance components through REML \citep{Wang1998, Ruppert2003}, which has been shown to be relatively robust to the misspecification of correlation structure \citep{Krivobokova2007}. Let \(\mathbf{R}^{\dagger}\) be the Moore-Penrose inverse of \(\mathbf{R}\), then a corresponding linear mixed effects model is given by \citep{Wang1998}
\[
\mathbf{Z}_{i} = \mathbf{B}\gamma_{i} + \boldsymbol{\epsilon}_{i}, \qquad
\gamma_{i} \sim N\left(\mathbf{0}, \frac{\sigma_{\epsilon,i}^{2}}{\lambda_{i}}\mathbf{R}^{\dagger}\right), \qquad
\boldsymbol{\epsilon}_{i} \sim N(\mathbf{0}, \sigma_{\epsilon,i}^{2}\mathbf{V}_{i}),
\]
where \(\gamma_{i}\) and \(\boldsymbol{\epsilon}_{i}\) are independent. Alternatively, let \(\mathbf{R} = Q^{T}P Q\), where \(Q\) is a \(K\times K\) invertible matrix, \(P = \operatorname{diag}(\mathbf{0}_{(K-r)}^{T}, \mathbf{1}_{r}^{T})\) and \(r = \mathrm{rank}(\mathbf{R})\), then a corresponding linear mixed effects model is given by
\[
\mathbf{Z}_{i} = \mathbf{B}Q^{-1}\begin{bmatrix} \mathbf{b}_{i} \\ \mathbf{u}_{i} \end{bmatrix} + \boldsymbol{\epsilon}_{i}, \qquad
\mathbf{u}_{i} \sim N\left(\mathbf{0}, \frac{\sigma_{\epsilon,i}^{2}}{\lambda_{i}}\mathbf{I}\right), \qquad
\boldsymbol{\epsilon}_{i} \sim N(\mathbf{0}, \sigma_{\epsilon,i}^{2}\mathbf{V}_{i}),
\]
where \(\begin{bmatrix} \mathbf{b}_{i} \\ \mathbf{u}_{i} \end{bmatrix} = Q\gamma_{i}\), \(\mathbf{b}_{i}\) is a fixed vector of length-\((K-r)\), \(\mathbf{u}_{i}\) is a random vector of length-\(r\), and \(\mathbf{u}_{i}\) and \(\boldsymbol{\epsilon}_{i}\) are independent.

When the error covariance matrix is the same for all subjects, a more parsimonious measurement error model is given by
\[
\mathbf{Z}_{i} = \mathbf{X}_{i} + \boldsymbol{\epsilon}_{i}, \qquad i = 1,\dots,n, \tag{16}
\]
where \(\boldsymbol{\epsilon}_{i}\) are normal with mean \(\mathbf{0}\) and variance \(\sigma_{\epsilon}^{2}\mathbf{V}\). We consider the following criterion to fit model (16), which is a natural extension of criterion (15),
\[
\sum_{i=1}^{n} (\mathbf{Z}_{i} - \mathbf{B}\gamma_{i})^{T}\mathbf{V}^{-1}(\mathbf{Z}_{i} - \mathbf{B}\gamma_{i}) + \lambda \sum_{i=1}^{n} \gamma_{i}^{T}\mathbf{R}\gamma_{i}. \tag{17}
\]
Let \(\mathbf{Z} = [Z_{i}(t_{d})]_{d=1,\dots,D; i=1,\dots,n}\) be a \(D\times n\) matrix, \(\boldsymbol{\Gamma} = [\gamma_{ki}]_{k=1,\dots,K; i=1,\dots,n}\) be a \(K\times n\) matrix, then criterion (17) can be rewritten as
\[
[\operatorname{vec}(\mathbf{Z}) - (\mathbf{I}_{n}\otimes \mathbf{B})\operatorname{vec}(\boldsymbol{\Gamma})]^{T} (\mathbf{I}_{n}\otimes \mathbf{V})^{-1}[\operatorname{vec}(\mathbf{Z}) - (\mathbf{I}_{n}\otimes \mathbf{B})\operatorname{vec}(\boldsymbol{\Gamma})] + \lambda \operatorname{vec}(\boldsymbol{\Gamma})^{T}(\mathbf{I}_{n}\otimes \mathbf{R})\operatorname{vec}(\boldsymbol{\Gamma}),
\]
which is in the form of \((\tilde{\mathbf{Z}} - \tilde{\mathbf{B}}\tilde{\boldsymbol{\Gamma}})^{T}\tilde{\mathbf{V}}^{-1}(\tilde{\mathbf{Z}} - \tilde{\mathbf{B}}\tilde{\boldsymbol{\Gamma}}) + \lambda \tilde{\boldsymbol{\Gamma}}^{T}\tilde{\mathbf{R}}\tilde{\boldsymbol{\Gamma}}\), with \(\tilde{\mathbf{Z}} = \operatorname{vec}(\mathbf{Z})\), \(\tilde{\mathbf{B}} = \mathbf{I}_{n}\otimes \mathbf{B}\), \(\tilde{\boldsymbol{\Gamma}} = \operatorname{vec}(\boldsymbol{\Gamma})\), \(\tilde{\mathbf{V}} = \mathbf{I}_{n}\otimes \mathbf{V}\), and \(\tilde{\mathbf{R}} = \mathbf{I}_{n}\otimes \mathbf{R}\). Thus, criterion (17) can be related to that for a linear mixed effects model, as done in the previous paragraph, and \(\sigma_{\epsilon}^{2}\) and \(\mathbf{V}\) can be estimated as variance components through REML.

The rest of the SIMEX algorithm is the same, except for adding pseudo errors with amount \(\theta\) based on the following equation
\[
\mathbf{Z}_{\theta,i} = \mathbf{Z}_{i} + \sqrt{\theta}\,\boldsymbol{\epsilon}_{i}, \qquad i = 1,\dots,n,
\]
where \(\boldsymbol{\epsilon}_{i}\) are normal with mean \(\mathbf{0}\) and variance \(\sigma_{\epsilon,i}^{2}\mathbf{V}_{i}\). Then we have
\[
\operatorname{Var}(\mathbf{Z}_{\theta,i} \mid \mathbf{X}_{i}) = (1+\theta)\operatorname{Var}(\mathbf{Z}_{i} \mid \mathbf{X}_{i}) = (1+\theta)\sigma_{\epsilon,i}^{2}\mathbf{V}_{i}.
\]
Let \(\theta = -1\), then \(\operatorname{Var}(\mathbf{Z}_{\theta,i} \mid \mathbf{X}_{i}) = \mathbf{0}\). That is, \(\theta = -1\) corresponds the case of zero measurement error.

\subsection{Repeated measurements}

If the curves are observed repeatedly at several sessions for each subject (but just one response \(Y_{i}\) is observed per subject), then we consider the following multilevel functional model for measurement error
\[
\mathbf{Z}_{i,j} = \mathbf{X}_{i} + \mathbf{U}_{i,j} + \boldsymbol{\epsilon}_{i,j}, \qquad i = 1,\dots,I, \quad j = 1,\dots,J_{i}, \tag{18}
\]
where \(\mathbf{Z}_{i,j} = (Z_{i,j}(t_{d}))_{d=1,\dots,D}^{T}\) is the observed curve for subject \(i\) at session \(j\), \(\mathbf{X}_{i} = (X_{i}(t_{d}))_{d=1,\dots,D}^{T}\) is the discretized subject-specific function, \(\mathbf{U}_{i,j} = (U_{i,j}(t_{d}))_{d=1,\dots,D}^{T}\) is the discretized session-specific functional deviation, \(\boldsymbol{\epsilon}_{i,j} = (\epsilon_{i,j}(t_{d}))_{d=1,\dots,D}^{T}\) is measurement error, \(I\) is the number of subjects, and \(J_{i}\) is the number of sessions for subject \(i\). We assume that \(\mathbf{U}_{i,j}\) is normal with mean \(\mathbf{0}\) and variance \(\boldsymbol{\Sigma}_{U}\), \(\boldsymbol{\epsilon}_{i,j}\) is normal with mean \(\mathbf{0}\) and variance \(\boldsymbol{\Sigma}_{\epsilon}\), and \(\mathbf{X}_{i}\), \(\mathbf{U}_{i,j}\) and \(\boldsymbol{\epsilon}_{i,j}\) are mutually uncorrelated. Here, the covariance structures \(\boldsymbol{\Sigma}_{U}\) and \(\boldsymbol{\Sigma}_{\epsilon}\) are left completely unspecified. Then we have
\[
\operatorname{Cov}(\mathbf{Z}_{i,j}, \mathbf{Z}_{i,j'}) = \boldsymbol{\Sigma}_{X} + (\boldsymbol{\Sigma}_{U} + \boldsymbol{\Sigma}_{\epsilon}) \mathbf{1}(j = j'). \tag{19}
\]
The matrix \(\boldsymbol{\Sigma}_{U} + \boldsymbol{\Sigma}_{\epsilon}\) may be estimated by viewing equation (19) as a regression model \citep{Greven2010}, or, alternatively, by the method of moments, as done in a FANOVA model \citep{Di2009}. Thus, without parametrically specifying the covariance structure for \(\mathbf{U}_{i,j} + \boldsymbol{\epsilon}_{i,j}\), which as a whole is regarded as ``noise'' in this context, we are able to separate the true function and the noise function, when there are repeated measurements made for each subject. Note that, to further identify \(\boldsymbol{\Sigma}_{U}\) and \(\boldsymbol{\Sigma}_{\epsilon}\) separately, we would need to parametrically specify the structure for \(\boldsymbol{\Sigma}_{\epsilon}\), and most commonly, it is assumed that \(\boldsymbol{\Sigma}_{\epsilon} = \sigma_{\epsilon}^{2}\mathbf{I}\).

We consider using the average observed curves by subject \(\bar{\mathbf{Z}}_{i,\cdot} = \mathbf{X}_{i} + \bar{\mathbf{U}}_{i,\cdot} + \bar{\boldsymbol{\epsilon}}_{i,\cdot}\) as a starting point to implement the SIMEX algorithm. We can add pseudo errors \(\boldsymbol{\tau}_{i}\) with amount \(\theta\) based on the following equation
\[
\bar{\mathbf{Z}}_{\theta,i,\cdot} = \bar{\mathbf{Z}}_{i,\cdot} + \sqrt{\theta}\,\boldsymbol{\tau}_{i}, \qquad i = 1,\dots,I,
\]
where \(\boldsymbol{\tau}_{i}\) is normal with mean \(\mathbf{0}\) and variance \(\boldsymbol{\Sigma}_{\tau,i} = (\boldsymbol{\Sigma}_{U} + \boldsymbol{\Sigma}_{\epsilon}) / J_{i}\). Then we have
\[
\operatorname{Var}(\bar{\mathbf{Z}}_{i,\cdot} \mid \mathbf{X}_{i}) = (\boldsymbol{\Sigma}_{U} + \boldsymbol{\Sigma}_{\epsilon}) / J_{i} = \boldsymbol{\Sigma}_{\tau,i},
\]
\[
\operatorname{Var}(\bar{\mathbf{Z}}_{\theta,i,\cdot} \mid \mathbf{X}_{i}) = (1+\theta)\operatorname{Var}(\bar{\mathbf{Z}}_{i,\cdot} \mid \mathbf{X}_{i}) = (1+\theta)\boldsymbol{\Sigma}_{\tau,i}.
\]
Let \(\theta = -1\), then \(\operatorname{Var}(\bar{\mathbf{Z}}_{\theta,i,\cdot} \mid \mathbf{X}_{i}) = \mathbf{0}\). That is, \(\theta = -1\) corresponds the case of zero measurement error.

\subsection{Generalized functional linear regression}

Suppose that the response \(Y_{i}\) has a general exponential family distribution with \(E(Y_{i}) = \mu_{i}\). Then we can consider a generalized functional linear model, given by
\[
g(\mu_{i}) = \beta_{0} + \int_{0}^{1} X_{i}(t)\beta(t) dt, \qquad i = 1,\dots,n,
\]
where \(g\) is a link function. The ideas for fitting generalized functional linear model resemble those for scalar-on-function model. The SIMEX algorithm extends easily, since the algorithm only requires that the coefficient function estimate \(\hat{\boldsymbol{\beta}}_{k}\) is a function of the curves with added error \(\mathbf{Z}_{\theta_{k},i}\) and the response \(Y_{i}\), \(i = 1,\dots,n\).

\subsection{Image predictors with measurement error}

In neuroimaging studies, it is of our interest to incorporate 2D- or 3D-brain imaging data to predict scalar clinical outcomes. The problem can be formulated as a generalized functional linear model
\[
g(\mu_{i}) = \beta_{0} + \int_{\mathcal{T}} X_{i}(\mathbf{t}) \beta(\mathbf{t}) d\mathbf{t}, \qquad i = 1,\dots,n,
\]
where \(X_{i}(\mathbf{t})\) is the image regressor, \(\beta(\mathbf{t})\) is the coefficient image, \(\mathbf{t} \in \mathbb{R}^{p}\) and \(\mathcal{T} \subset \mathbb{R}^{p}\) (most commonly, \(p = 2\) or 3). When the images are observed with measurement error, a measurement error model is given by
\[
\mathbf{Z}_{i} = \mathbf{X}_{i} + \boldsymbol{\epsilon}_{i},
\]
where \(\mathbf{Z}_{i} = (Z_{i}(\mathbf{t}_{1}), \dots, Z_{i}(\mathbf{t}_{D}))^{T}\) is the observed image at spatial locations \(\mathbf{t}_{1},\dots,\mathbf{t}_{D}\), \(\mathbf{X}_{i} = (X_{i}(\mathbf{t}_{1}), \dots, X_{i}(\mathbf{t}_{D}))^{T}\) is the (discretized) true image, and measurement error \(\boldsymbol{\epsilon}_{i} = (\epsilon_{id})_{d=1,\dots,D}^{T}\) is assumed to be iid normal with mean \(\mathbf{0}\) and variance \(\sigma_{\epsilon}^{2}\mathbf{I}\), for simplicity. Several methods have been developed for generalized functional linear model with image as predictors, including \citet{ReissOgden2010} and \citet{Goldsmith2014}, and may be potentially applied together with the SIMEX algorithm to handle measurement error in the observed images, which is certainly an issue with noisy images such as those obtained in brain imaging studies. The error variance \(\sigma_{\epsilon}^{2}\) may be estimated via thin plate (regression) splines or tensor product splines \citep[see pp.~167-168]{Wood2006a}. For a given added amount of measurement error \(\theta_{k}\), the corresponding coefficient image estimate \(\hat{\boldsymbol{\beta}}_{k}(\mathbf{t})\) can be obtained by Monte Carlo simulation. The number of simulations \(B\) should be greater compared to the 1-D case, since higher dimensional integration is generally more time consuming.

The extrapolation methods must be extended from modeling a sequence of 1D coefficient functions to modeling a sequence of 2D or 3D coefficient images, with respect to increasing amounts of added error. Suppose that the dimension of image \(p = 2\), for ease of presentation. For linear extrapolation, an image-on-scalar regression model is introduced
\[
E(\hat{\boldsymbol{\beta}}_{k}(\mathbf{t})) = b_{0}(\mathbf{t}) + b_{1}(\mathbf{t})\theta_{k}, \qquad k = 1,\dots,K. \tag{20}
\]
Here, \(\theta_{k}\) may be easily generalized to other functional forms. Model (20) may be fitted by minimizing the following penalized least squares criterion
\[
\sum_{k=1}^{K} \int_{\mathcal{T}} \{\beta_{k}(\mathbf{t}) - b_{0}(\mathbf{t}) - b_{1}(\mathbf{t})\theta_{k}\}^{2} d\mathbf{t} + \lambda \{ J(b_{0}) + J(b_{1}) \}, \tag{21}
\]
where \(J\) is a penalty functional given by the thin plate penalty
\[
J(f) = \int\int \left\{ \left(\frac{\partial^{2}f}{\partial x_{1}^{2}}\right)^{2} + 2\left(\frac{\partial^{2}f}{\partial x_{1}\partial x_{2}}\right)^{2} + \left(\frac{\partial^{2}f}{\partial x_{2}^{2}}\right)^{2} \right\} dx_{1}dx_{2}.
\]
Let \(\hat{\boldsymbol{\beta}}(\mathbf{t}) = (\hat{\beta}_{k}(\mathbf{t}))_{k=1,\dots,K}^{T}\), \(\mathbf{b}(\mathbf{t}) = (b_{m}(\mathbf{t}))_{m=0,1}^{T}\) and \(\boldsymbol{\Theta} = [\mathbf{1}, (\theta_{k})_{k=1,\dots,K}^{T}]\). Consider representing \(\hat{\boldsymbol{\beta}}(\mathbf{t})\) and \(\mathbf{b}(\mathbf{t})\) by tensor product spline basis functions, i.e., \(\hat{\boldsymbol{\beta}}(\mathbf{t}) = \mathbf{C}\mathbf{B}(\mathbf{t})\) and \(\mathbf{b}(\mathbf{t}) = \mathbf{A}\mathbf{B}(\mathbf{t})\), where \(\mathbf{B}(\mathbf{t}) = (B_{1}\otimes B_{2})(\mathbf{t})\) is a vector of tensor product spline basis functions evaluated at \(\mathbf{t}\), and \(\mathbf{C}\) and \(\mathbf{A}\) are matrices of coefficients for \(\hat{\boldsymbol{\beta}}(\mathbf{t})\) and \(\mathbf{b}(\mathbf{t})\), respectively. Then, the penalty term \(\lambda \{ J(b_{0}) + J(b_{1}) \}\) in criterion (21) can be rewritten as
\[
\lambda \operatorname{vec}(\mathbf{A}^{T})^{T} (\mathbf{I}_{2}\otimes \mathbf{R}) \operatorname{vec}(\mathbf{A}^{T}), \tag{22}
\]
where \(\mathbf{R}\) is given by
\[
\mathbf{R} = \int\int \left\{ \frac{\partial^{2}\mathbf{B}}{\partial t_{1}^{2}} \frac{\partial^{2}\mathbf{B}^{T}}{\partial t_{1}^{2}} + 2 \frac{\partial^{2}\mathbf{B}}{\partial t_{1}\partial t_{2}} \frac{\partial^{2}\mathbf{B}^{T}}{\partial t_{1}\partial t_{2}} + \frac{\partial^{2}\mathbf{B}}{\partial t_{2}^{2}} \frac{\partial^{2}\mathbf{B}^{T}}{\partial t_{2}^{2}} \right\} dt_{1}dt_{2}.
\]
Following derivations as in \citet{ReissHuang2010}, the first term \(\sum_{k=1}^{K} \int_{\mathcal{T}} \{\beta_{k}(\mathbf{t}) - b_{0}(\mathbf{t}) - b_{1}(\mathbf{t})\theta_{k}\}^{2} d\mathbf{t}\) in criterion (21) can be rewritten as
\[
\| \operatorname{vec}(\mathbf{J}_{BB}^{1/2}\mathbf{C}^{T}) - (\boldsymbol{\Theta}\otimes \mathbf{J}_{BB}^{1/2})\operatorname{vec}(\mathbf{A}^{T}) \|^{2}, \tag{23}
\]
where \(\mathbf{J}_{BB} = \int_{\mathcal{T}} \mathbf{B}(\mathbf{t})\mathbf{B}(\mathbf{t})^{T} d\mathbf{t}\). Then, criterion (21), which is the sum of the two terms (23) and (22), can be viewed as a criterion for generalized ridge regression, whose minimizer can be obtained easily. The tuning parameter \(\lambda\) may be chosen by leave-one-image-out CV.

For nonlinear extrapolation, the following model may be considered
\[
E(\hat{\boldsymbol{\beta}}) = f(t_{1}, t_{2}, \theta),
\]
where \(f\) is a smooth function in \(\mathbb{R}^{3}\). The model can be fitted by thin plate splines \citep{GreenSilverman1994} or tensor product splines \citep{Wood2006a}.

For local polynomial extrapolation, we consider fitting a localized model at \(\theta_{0} = -1\) and controlling the smoothness with respect to \(\mathbf{t}\), by minimizing the following penalized criterion
\[
\sum_{k=1}^{K} \int_{\mathcal{T}} \{\beta_{k}(\mathbf{t}) - \sum_{m=0}^{M} b_{m}(\mathbf{t})(\theta_{0} - \theta_{k})^{m}\}^{2} K_{h}(\theta_{0} - \theta_{k}) d\mathbf{t} + \lambda \sum_{m=0}^{M} J(b_{m}), \tag{24}
\]
where \(J\) is the thin plate penalty. Criterion (24) may be transformed to a criterion for the image-on-scalar regression model, similar to what have been done in Section~3.2.3. The bandwidth \(h\) and the tuning parameter \(\lambda\) may be chosen by leave-one-image-out CV.

 \section{Application to diffusion tensor imaging data}\label{application}

\subsection{Background}

Multiple sclerosis (MS) is a disabling neurological disease that occurs mainly in the white matter of the central nervous system \citep{GoldbergZimring2005}. The corpus callosum is the largest area of white matter in the brain and it connects the two cerebral hemispheres. It is usually abnormal for MS cases \citep{Ozturk2010}. Diffusion tensor imaging (DTI), a brain imaging technique which takes advantage of diffusion anisotropy, can reveal abnormality in the white matter and is widely used in the study of MS \citep{GoldbergZimring2005}. A primary measurement derived from DTI is called fractional anisotropy (FA), which reflects the relative anisotropy of diffusion direction. It was found to decrease in the corpus callosum for MS cases in previous studies \citep{Ozturk2010}.

The DTI data we analyze were collected at Johns Hopkins University and the Kennedy-Krieger Institute and consist of FA tract profiles of the corpus callosum for both MS cases and controls at multiple visits. The dataset is available in the refund package in R \citep{Crainiceanu2014}. We will focus on the classification of MS case status (1 for case, 0 for control) using the FA tract profiles of the corpus callosum. There are 100 (70\%) MS cases and 42 (30\%) controls, and each subject is measured once for each visit. There are up to 8 visits per subject, giving 382 visits in total. The FA tract profiles of corpus callosum are sampled at 93 positions. Six (2\%) of the 382 visits have missing values in the profiles. After discarding profiles with missing values, the number of MS cases and controls remains unchanged. The number of visits ranges from 1 to 7, with a median of 2.

\begin{figure}[tbp]
\centering
\includegraphics[width=0.6\linewidth]{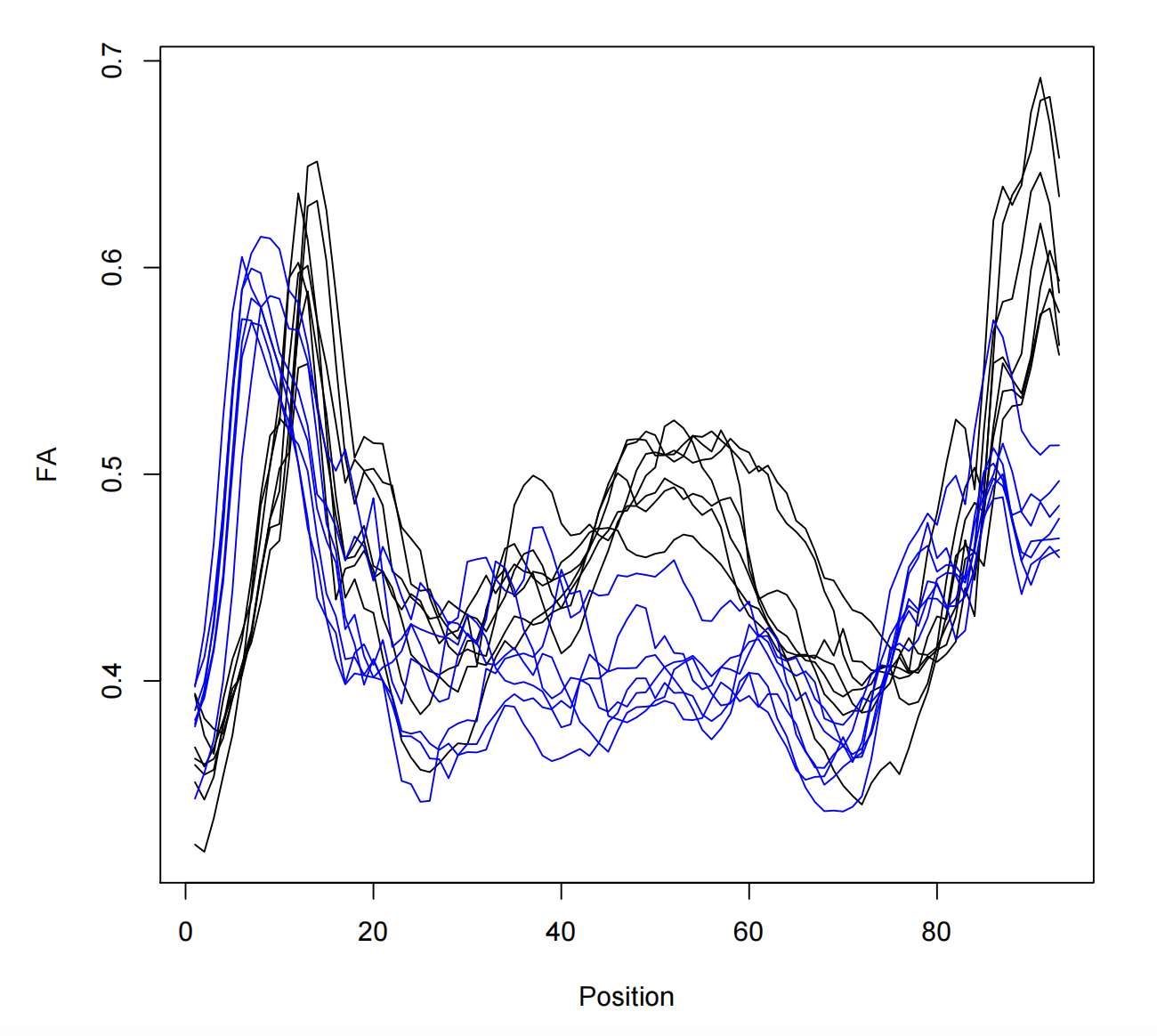}
\caption{The FA tract profiles for two MS cases measured at all visits. Two subjects are indicated by black and blue, respectively.}
\label{fig:profiles}
\end{figure}

As an example, Figure~\ref{fig:profiles} shows the FA tract profiles of corpus callosum for two subjects, both MS cases, measured at all visits. The profiles have variation between subjects and between visits and are subject to measurement error. To use the FA tract profiles of corpus callosum as functional regressors to classify MS case status, we need to properly account for the effect caused by between-visit variation and measurement error. SIMEX is a method that we described to deal with measurement error in the functional regressors for scalar-on-function regression. It can be extended to handle multilevel functional measurements and the response variable that has an exponential family distribution. In this section, we will apply the SIMEX algorithm for generalized functional linear regression where the functional regressors have multilevel measurements to the DTI dataset. We will first estimate between-visit and measurement error variance of the FA tract profiles of corpus callosum, and then build generalized functional regression model to classify MS case status based on the profiles, accounting for both between-visit variation and measurement error.

\subsection{Variance decomposition}

\begin{figure}[htbp]
\centering
\includegraphics[width=\linewidth]{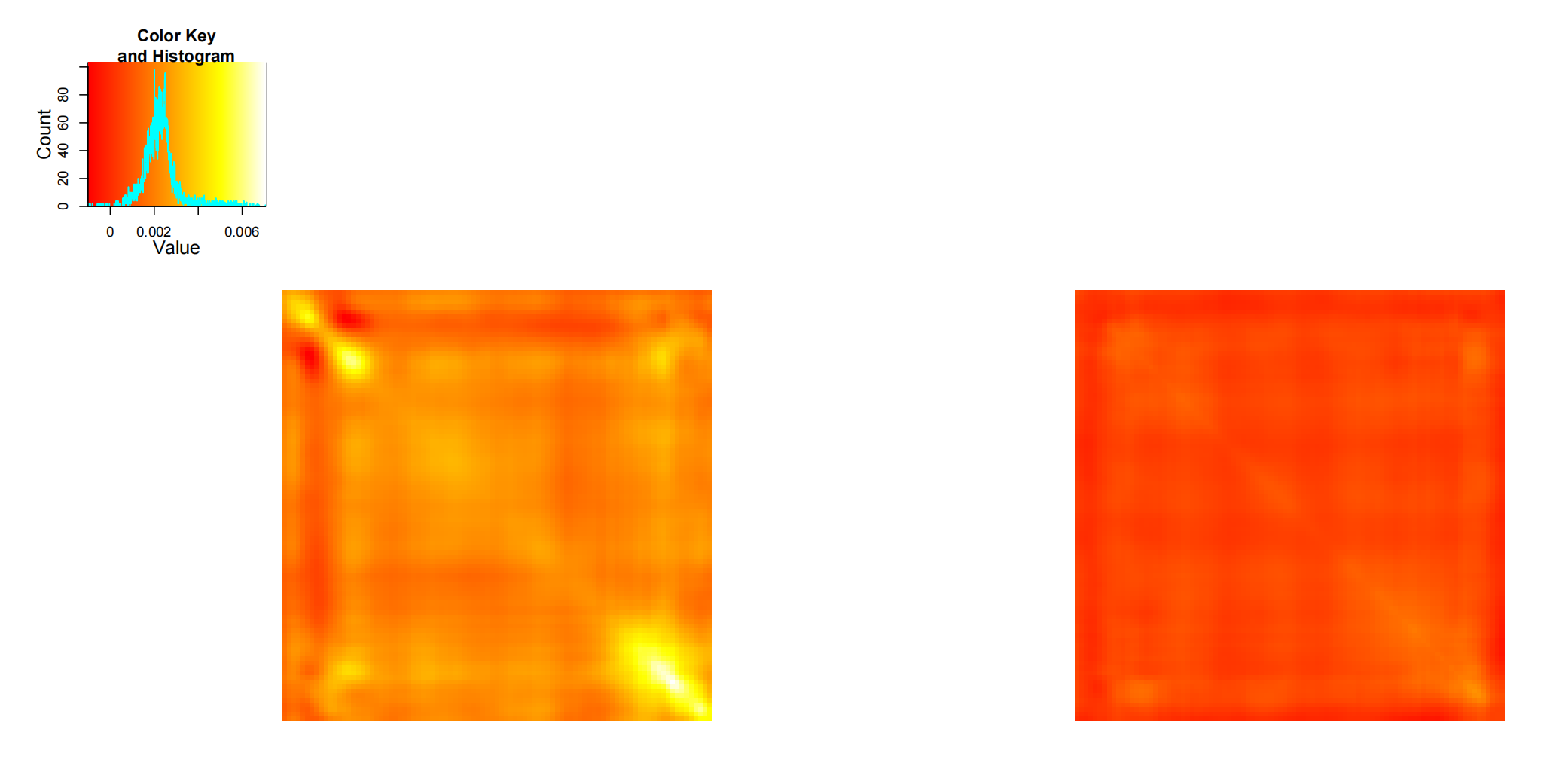}
\caption{Heat maps of the estimated covariance matrix for the subject level (left) and the estimated covariance matrix for the visit level plus measurement error (right). Upper left: color key and histogram of heat maps.}
\label{fig:cov}
\end{figure}

As illustrated in Section~7.1, the FA tract profiles of corpus callosum have between-subject and between-visit variation and are subject to measurement error. Thus, we consider the multilevel functional model (18) to characterize variation at different levels. Regarding equation (19) as a linear regression model, the covariance matrix for the subject level \(\boldsymbol{\Sigma}_{X}\) and the covariance matrix for the visit level plus measurement error \(\boldsymbol{\Sigma}_{U} + \boldsymbol{\Sigma}_{\epsilon}\) can be estimated using BLUE and the estimates are shown in Figure~\ref{fig:cov}. Note that the estimated covariance matrices are not guaranteed to be nonnegative definite. A simple correction is to perform eigendecomposition on the estimated covariance matrices and then reconstruct the matrices using only the eigenvectors that have positive eigenvalues. The total proportion of variance explained by the visit level and measurement error together is
\[
\frac{\operatorname{trace}(\hat{\boldsymbol{\Sigma}}_{U} + \hat{\boldsymbol{\Sigma}}_{\epsilon})}{\operatorname{trace}(\hat{\boldsymbol{\Sigma}}_{X}) + \operatorname{trace}(\hat{\boldsymbol{\Sigma}}_{U} + \hat{\boldsymbol{\Sigma}}_{\epsilon})} = 23\%.
\]

\subsection{Coefficient function estimates}

Ideally, if we can observe \(\{X_{i}(t), Y_{i}\}_{i=1,\dots,I}\), where \(X_{i}(t)\) is the true FA tract profile of corpus callosum, \(Y_{i}\) is the MS case status (1 for case, 0 for control), and \(i\) is the index for subjects 1 through \(I\), then we can apply the following functional logistic regression model
\[
\log \frac{P(Y_{i} = 1)}{1 - P(Y_{i} = 1)} = \beta_{0} + \int_{0}^{1} X_{i}(t)\beta(t) dt, \qquad i = 1,\dots,I. \tag{25}
\]
where \(\beta_{0}\) is the intercept and \(\beta(t)\) is the coefficient function. However, as described in the previous paragraph, the profiles are observed with between-visit variation and measurement error, which explains \(23\%\) of the total variance. One could average the profiles at all visits for each subject, but doing so may not effectively eliminate the effect caused by between-visit variation and measurement error, especially when more than half of the subjects only have 1 or 2 visits. To properly account for the effect caused by between-visit variation and measurement error, we apply the SIMEX algorithm described in Section~6.3.

The average profiles by subject may be used as a starting point to implement SIMEX. Let \(\bar{\mathbf{Z}}_{i,\cdot}\) be the average profile for subject \(i\). Conditional on the true profile \(\mathbf{X}_{i}\), \(\operatorname{Var}(\bar{\mathbf{Z}}_{i,\cdot} \mid \mathbf{X}_{i}) = (\boldsymbol{\Sigma}_{U} + \boldsymbol{\Sigma}_{\epsilon}) / J_{i}\), where \(J_{i}\) is the number of visits for subject \(i\). Thus we consider adding additional errors \(\boldsymbol{\tau}_{i}\) based on \(\bar{\mathbf{Z}}_{\theta,i,\cdot} = \bar{\mathbf{Z}}_{i,\cdot} + \sqrt{\theta}\,\boldsymbol{\tau}_{i}\), for \(i = 1,\dots,I\), where \(\boldsymbol{\tau}_{i}\) is normal with mean \(\mathbf{0}\) and variance \((\hat{\boldsymbol{\Sigma}}_{U} + \hat{\boldsymbol{\Sigma}}_{\epsilon}) / J_{i}\) and \(\theta\) controls for the amount of added errors. We repeat this simulation step \(B = 100\) times and choose \(\theta\) to be from 0 to 2.25 at intervals of 0.25. The functional logistic regression model is estimated by techniques presented in \citet{ReissOgden2010}. The tuning parameter and the number of B-spline basis functions are fixed for each \(\theta\) and each simulation. Estimates for the intercept \(\hat{\beta}_{\theta,0}\) and the coefficient function \(\hat{\beta}_{\theta}(t)\) for each \(\theta\) are obtained by taking the average over \(B\) simulations, respectively. Then the sequence of estimates is modeled as a function of \(\theta\) and extrapolated to where \(\theta = -1\), the case of zero-error. The extrapolation is done in three ways: linear, nonlinear and local polynomial. For the linear extrapolation method, the intercept estimate is obtained based on a simple linear regression model with \(\hat{\beta}_{\theta,0}\) as response and \(\theta\) as covariate, and the coefficient function estimate is obtained based on a function-on-scalar regression model as described in Section~3.2.1. For the nonlinear extrapolation method, the intercept estimate is obtained based on a smoothing splines model, where \(\hat{\beta}_{\theta,0}\) is modeled as a smooth function of \(\theta\), and the coefficient function estimate is obtained based on a tensor product spline model as described in Section~3.2.2. For the local polynomial extrapolation method, the intercept estimate is obtained based on a local quadratic regression model, and the coefficient function estimate is obtained based on (10) as described in Section~3.2.3, where the order of polynomial is quadratic.

\begin{figure}[tb]
\centering
\includegraphics[width=\linewidth]{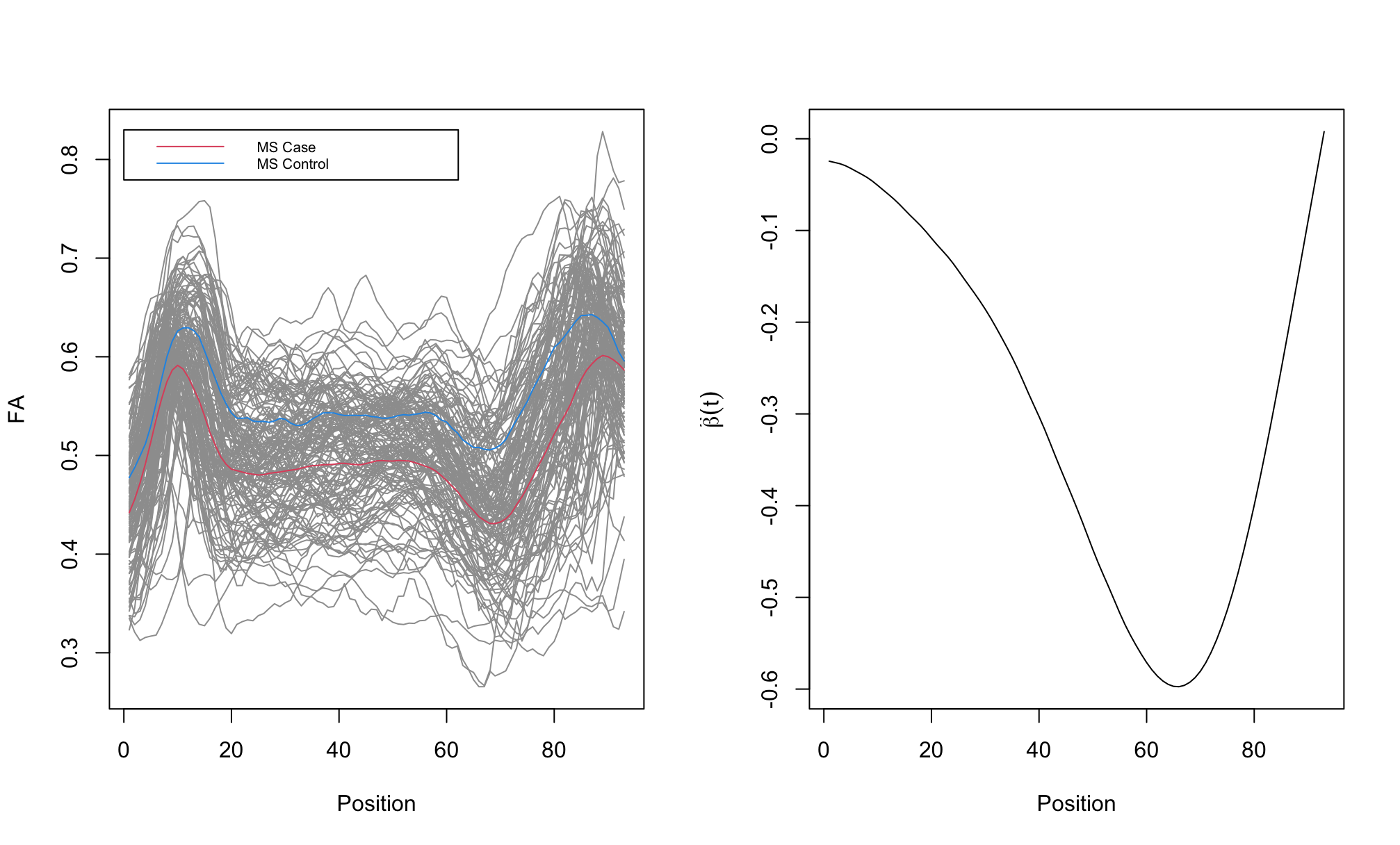}
\caption{Left: the average profiles by subject are indicated by grey, the average profile for MS cases are indicated by red, and the average profile for MS controls are indicated by blue. Right: the naive estimate of the coefficient function obtained by regressing the MS case status on the average profiles by subject.}
\label{fig:avgprof}
\end{figure}

\begin{figure}[tb]
\centering
\includegraphics[width=\linewidth]{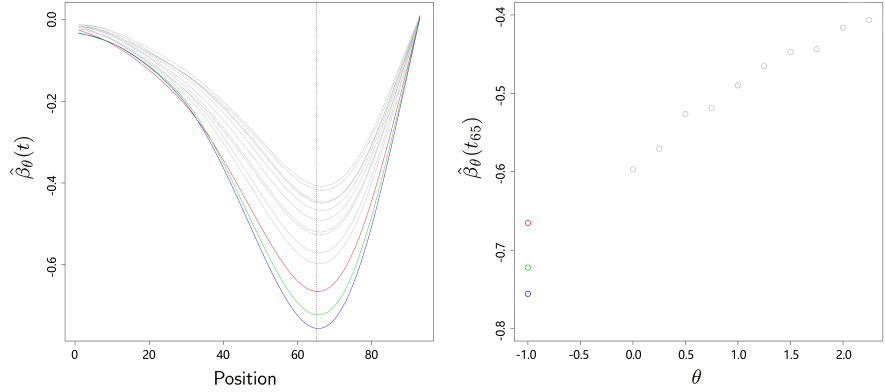}
\caption{Left: the sequence of estimates for the coefficient function $\hat{\beta}_{\theta}(t)$, for $\theta = 0, 0.25, \dots, 2.25$, are indicated by grey; the estimates obtained by linear, nonlinear and local polynomial extrapolation are indicated by red, green and blue, respectively; position 65 on the tract is indicated by the dashed line. Right: $\hat{\beta}_{\theta}(t_{65})$ against $\theta$ is indicated by grey; the estimates obtained by linear, nonlinear and local polynomial extrapolation are plotted at $\theta = -1$ and indicated by red, green and blue, respectively.}
\label{fig:coef_est}
\end{figure}

\begin{figure}[htbp]
\centering
\includegraphics[width= 0.6 \linewidth]{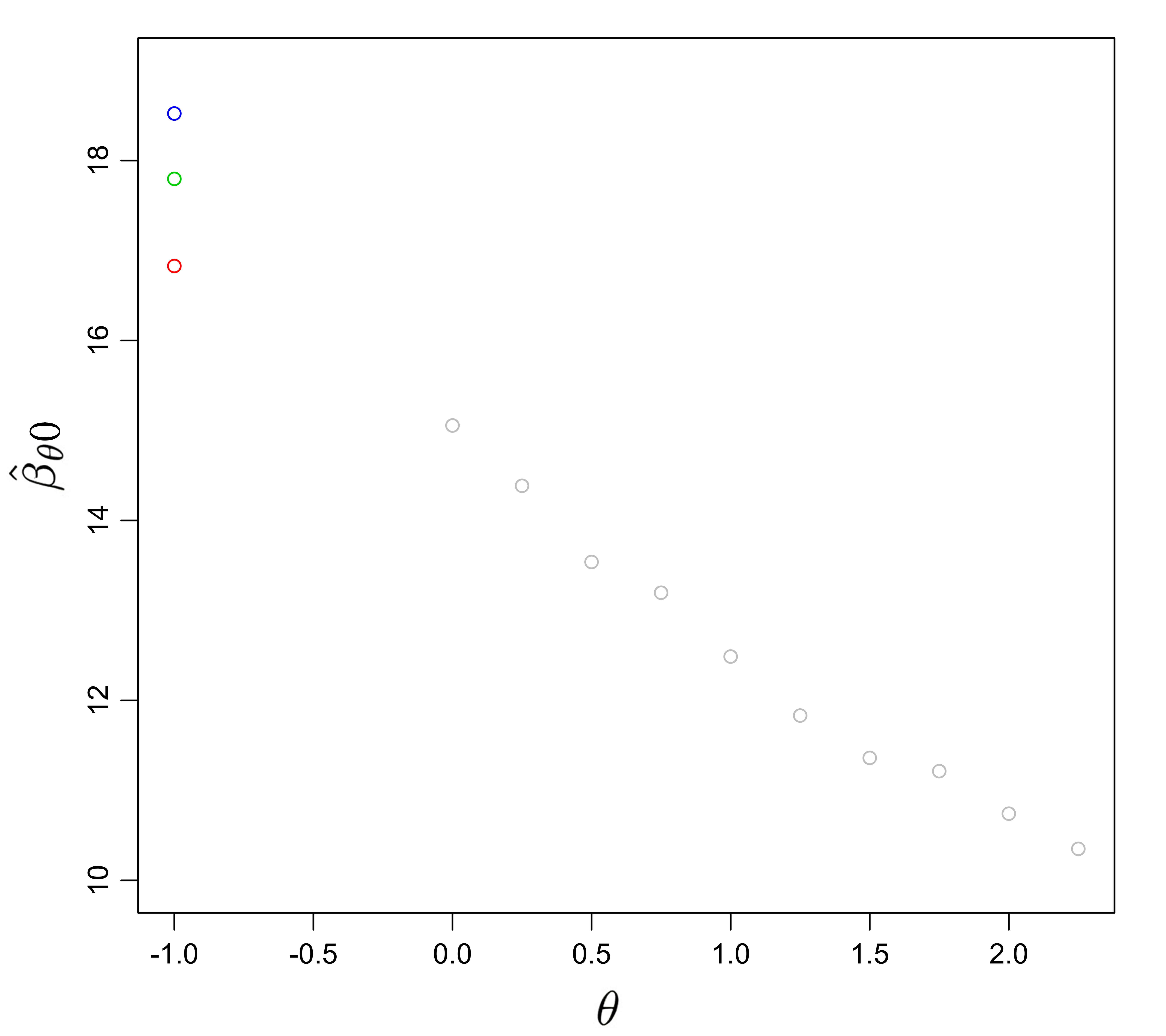}
\caption{The sequence of estimates for the intercept $\hat{\beta}_{\theta,0}$, for $\theta = 0, 0.25, \dots, 2.25$, are indicated by grey; the estimates obtained by linear, nonlinear and local polynomial extrapolation are plotted at $\theta = -1$ and indicated by red, green and blue, respectively.}
\label{fig:intercept}
\end{figure}

As a comparison, the result of regressing MS case status \(Y_{i}\) on the average profiles \(\bar{\mathbf{Z}}_{i,\cdot}\), which we refer as the naive method, is presented. Figure~\ref{fig:avgprof} shows the average profiles by MS case status and the naive estimate of the coefficient function. We can see that MS cases tend to have lower FA for corpus callosum compared to controls and the difference is most substantial at around position 65 on the tract (the confidence interval will be presented in the next paragraph). Figure~\ref{fig:coef_est} shows the estimates of the coefficient function by SIMEX with linear (red), nonlinear (green) and local polynomial (blue) extrapolation. To see which extrapolation method uses a model that performs the best, we calculate the leave-one-curve-out CV score for each extrapolation method. The CV scores for linear, nonlinear and local polynomial extrapolation are 0.0063, 0.0034, and 0.0022, respectively. Based on this criterion, the model used by local polynomial extrapolation performs better than that used by nonlinear extrapolation, followed by linear extrapolation. Figure~\ref{fig:intercept} shows the intercept estimates by the naive method (corresponding to \(\theta = 0\)) and by SIMEX with linear (red), nonlinear (green) and local polynomial (blue) extrapolation.

\subsection{Bootstrapping SIMEX}

The SIMEX algorithm is relatively simple to implement, but it is not straightforward to derive the asymptotic distribution for the estimators \citep{CookStefanski1994}. In this section, we will use the bootstrap to calculate the pointwise confidence intervals for the SIMEX method and compare to that for the naive method.

There are several resampling schemes available for bootstrapping a regression model \citep{EfronTibshirani1994, DavisonHinkley1997}. For a classic (generalized) linear regression model, bootstrap may be done completely nonparametrically by resampling the pairs of the response and the covariates. Alternatively, it may be done via a model-based resampling scheme: conditional on the covariates, the bootstrapped response is generated by adding a random error, resampled from the centered residuals, to the fitted value from the model, or generated by resampling from the fitted distributional model. The residual bootstrap for generalized linear regression model involves resampling from the Pearson residuals or the deviance residuals, but one major drawback is that it can generate non-integer responses, while the actual responses should be integer-valued \citep[see pp.~333-334]{DavisonHinkley1997}.

Moving forward to functional (generalized) linear regression, most of the literature to date focuses on adapting the resampling schemes available for bootstrapping classic regression models to the functional situation. \citet{James2009} used the nonparametric bootstrap for functional linear regression in an application to Canadian weather data. \citet{ReissOgden2010} constructed simultaneous confidence bands by the nonparametric bootstrap for functional generalized linear regression. Some research work has also been done to justify the asymptotic validity of bootstrapping functional regression models. \citet{GonzalezManteiga2011} showed the validity of the residual bootstrap for FPCA-type estimates in functional linear models. For the point impact model in functional linear regression, \citet{McKeagueSen2010} showed the consistency of the residual bootstrap, whereas the nonparametric bootstrap is not consistent. The latter result suggests more careful choice of the resampling scheme has to be made when bootstrapping functional regression models. In light of that, we will use the model-based bootstrap in our case of functional logistic regression. Specifically, we will resample from the fitted Bernoulli model, since the residual bootstrap might result in bootstrapped responses other than 0 or 1.

\begin{figure}[htbp]
\centering
\includegraphics[width=\linewidth]{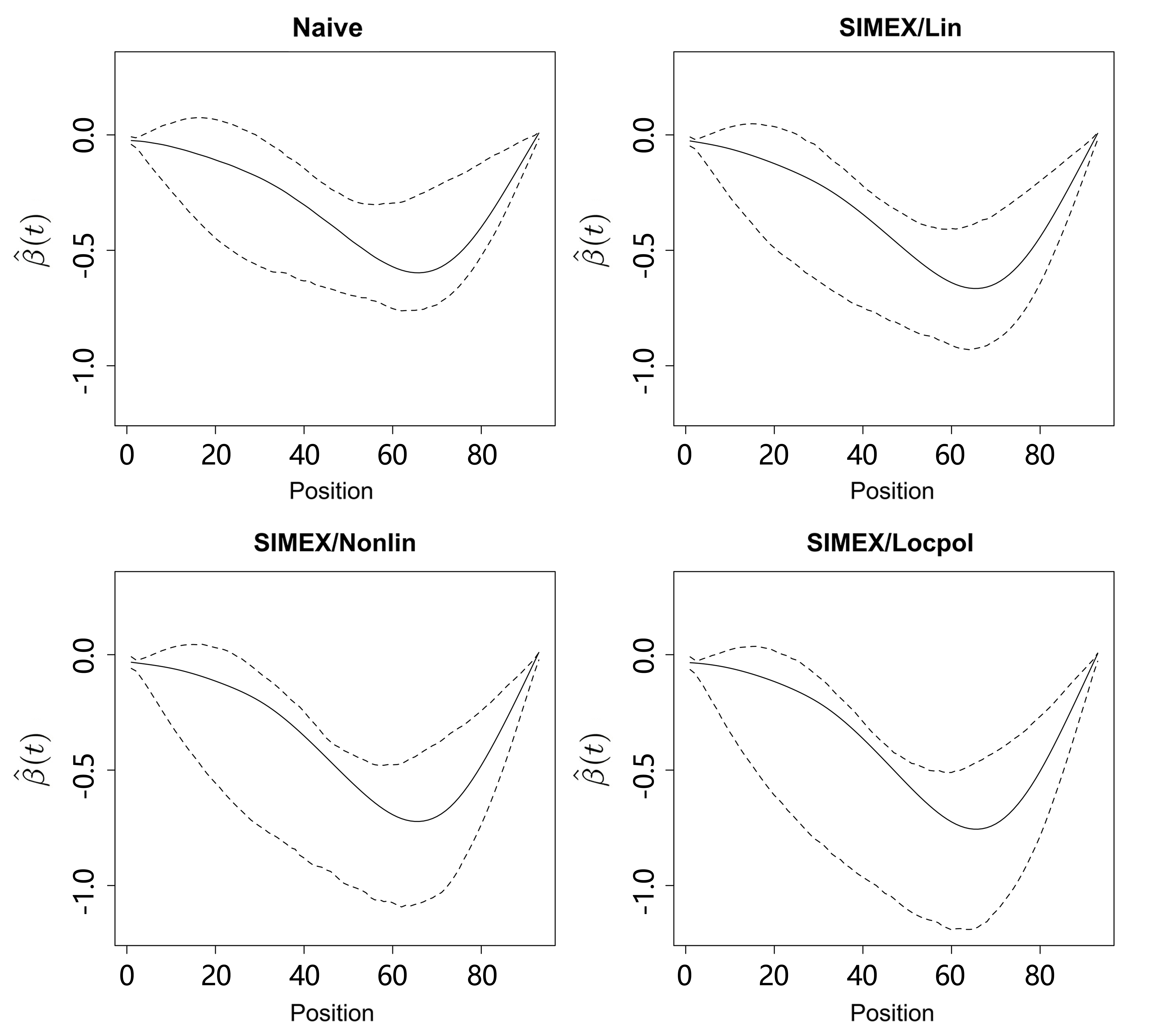}
\caption{The estimated coefficient functions (solid line) and 95\% pointwise confidence intervals (dashed line) by naive (top left) and SIMEX with linear (top right), nonlinear (bottom left) and local polynomial (bottom right) extrapolation.}
\label{fig:ci}
\end{figure}

Figure~\ref{fig:ci} shows the 95\% pointwise confidence intervals obtained via bootstrap for the naive method and the SIMEX methods with linear, nonlinear, and local polynomial extrapolation. The bootstrap sampling is repeated 1000 times. The tuning parameter and the number of B-spline basis functions used in fitting the functional logistic regression model are kept fixed across the bootstrapped samples. The pointwise confidence interval for SIMEX with local polynomial extrapolation tends to be wider than that for nonlinear, followed by linear and then naive, which might suggest the bias-variance tradeoff.

\section{Discussion}\label{discussion}

In the simulation study, we find that for relatively large measurement error, SIMEX extrapolation methods tend to be conservative. This might be because the attenuation effect caused by the measurement error is large, so that the sequence of coefficient function estimates corresponding to added error cannot sufficiently reflect the curvature along \(\theta\) near the case of zero-error. SIMEX might be potentially applied together with most estimation methods for scalar-on-function regression model, including \citet{Cardot2003}, \citet{ReissOgden2007}, \citet{Goldsmith2011a}, which could have a substantial impact on the performance of SIMEX. In practice, these estimation methods sometimes provide quite different coefficient function estimates. We need to choose the estimation method carefully according to the dataset before implementing SIMEX. In addition, in the simulation study, we find that pre-smoothing with smoothing splines performs similarly to the naive method, because the naive method, implemented as in \citet{Cardot2003}, inherently involves projecting the observed curves onto a B-spline basis, which may be understood as ``smoothing by splines'' as well.

\begin{figure}[tb]
\centering
\includegraphics[width=0.6\linewidth]{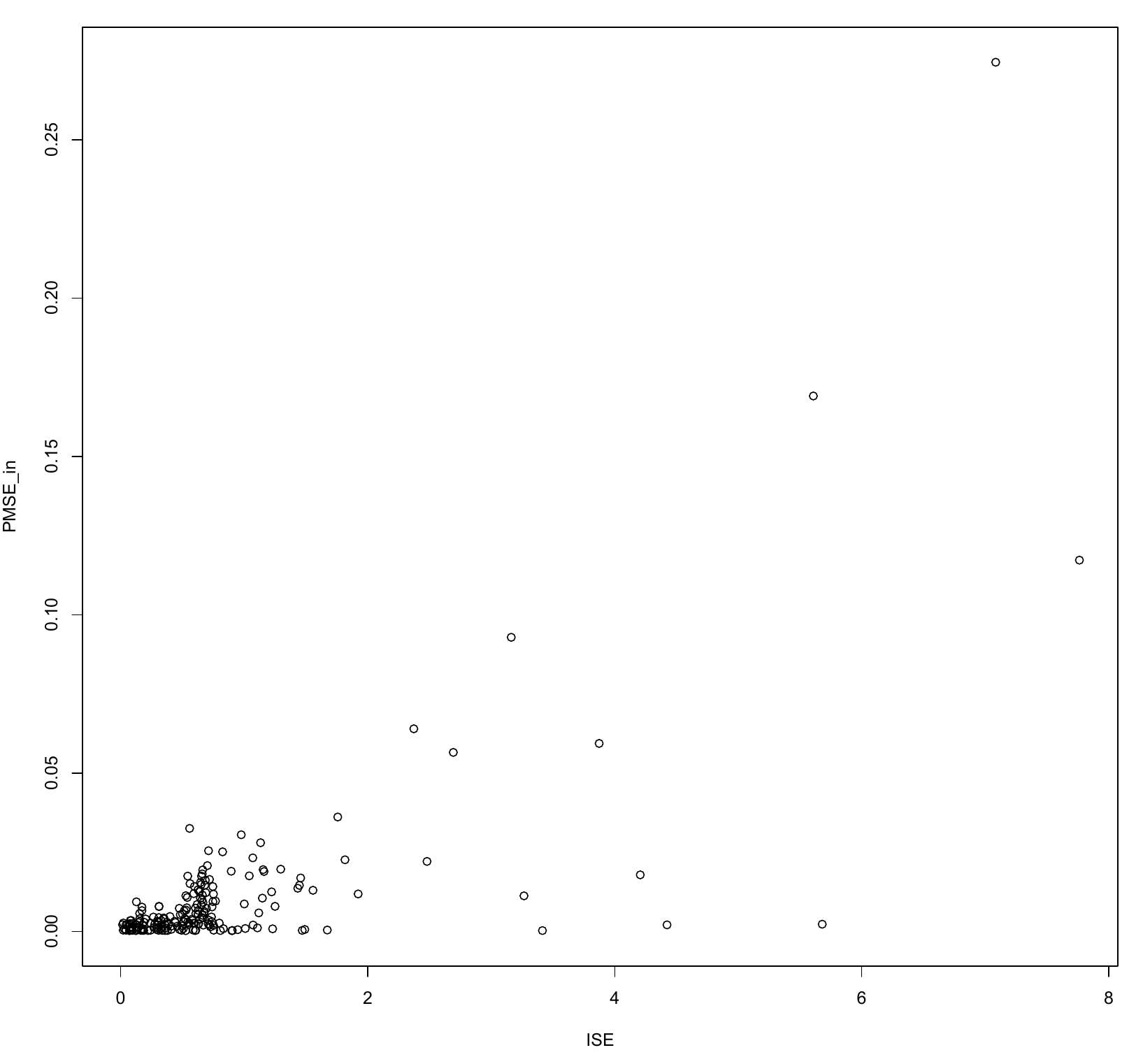} 
\caption{$\mathrm{PMSEin}(Y)$ vs. $\mathrm{ISE}(\beta)$ for pre-smoothing with FPCA when $\sigma_{\epsilon} = 3$.}
\label{fig:pmse_ise}
\end{figure}

\begin{figure}[htbp]
\centering
\includegraphics[width=\linewidth]{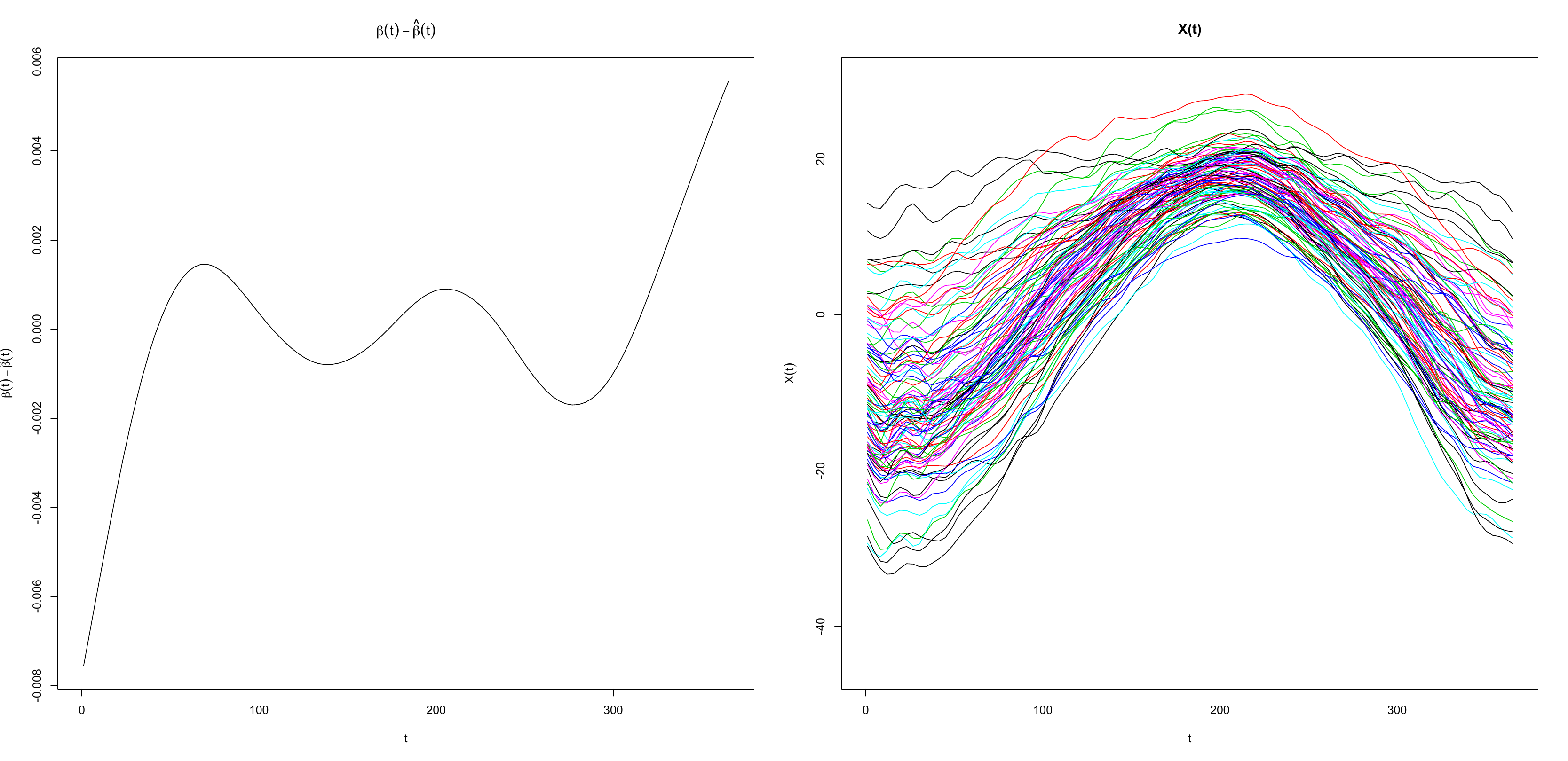}
\caption{These plots illustrate one sample where the difference between $\hat{\beta}$ and $\beta$ is approximately orthogonal to $X_{i}$, for $i = 1,\dots,n$, for pre-smoothing with FPCA when $\sigma_{\epsilon} = 3$. Left: the plot of $\beta - \hat{\beta}$. Right: the plot of the true curves $X_{i}$.}
\label{fig:orthogonal}
\end{figure}

Regarding the observation that pre-smoothing with FPCA does not estimate well but predicts well in the simulation study, we give a possible explanation as follows. Consider model (1) and suppose that \(\beta^{*}(t)\) is any function which is orthogonal to functional regressors \(X_{i}(t)\), for \(i = 1,\dots,n\), then adding \(\beta^{*}(t)\) to the true coefficient function \(\beta(t)\) will not affect the prediction. Figure~\ref{fig:pmse_ise} shows the plot of \(\mathrm{PMSEin}(Y)\) against \(\mathrm{ISE}(\beta)\) for pre-smoothing with FPCA, when the standard deviation of measurement error \(\sigma_{\epsilon}\) is 3. We can see that there are quite a few samples, for which \(\mathrm{ISE}(\beta)\) is very large but \(\mathrm{PMSEin}(Y)\) is quite small. We find that for those samples, the difference between \(\hat{\beta}\) and \(\beta\) is approximately orthogonal to \(X_{i}\), for \(i = 1,\dots,n\). Figure~\ref{fig:orthogonal} illustrates one of those samples. A similar problem of nonidentifiability has been reported in \citet{Cardot2003}, where they found that a few ``outliers'' resulted in a high variance in estimating the coefficient function. \citet{ScheiplGreven2012} also described the nonidentifiability problem in the context of penalized function-on-function regression model. When the ``effective'' number of principal components of functional regressors (the smallest number of principal components which explains more than, for example, 99\% of the total variance) is very small, particularly, smaller than the number of basis functions used to expand the coefficient function, identifiability is likely to be an issue. In this simulation, the curves smoothed by FPCA require only 3 or 4 ``effective'' number of principal components, smaller than 14, the number of B-spline basis functions. Such problem of nonidentifiability, however, is supposed to be solved by adding the roughness penalty, where the penalty matrix is nonsingular. But we find that for those samples in which the nonidentifiability problem might occur, the chosen tuning parameter \(\rho\) in the scalar-on-function regression model is often too small. This might relate to the undersmoothing behavior of GCV \citep{WahbaWang1995, Cardot2003}. If, in the simulation setup, the true curves \(\mathbf{X}_{i}\) were generated by a larger ``effective'' number of principal components (the sample size remains the same), we find that the nonidentifiability problem might become less likely to occur, but the method does not estimate or predict very well. This might be because FPCA does not perform as well in uncovering the true curves, when those are not dominated by only a few number of dimensions.

In future research, we will further investigate the asymptotic behavior of the SIMEX estimator. We will also investigate if SIMEX is robust to non-normal measurement error. In addition, we are interested in incorporating binary functional data as predictors in generalized functional linear model and developing methods to deal with misclassification in binary functional predictors.

\section*{Supplementary Materials}

\subsection*{SIMEX Algorithm for Scalar-on-function Regression}

\begin{enumerate}
\item Based on model (2), estimate the error variance \(\sigma_{\epsilon}^{2}\) via smoothing methods.
\item For each \(\theta_{k} \in \{\theta_{1}, \dots, \theta_{K}\}\) and for each \(b \in \{1, \dots, B\}\):
\begin{enumerate}
\item By Monte Carlo simulation, add an additional amount of measurement error, where the amount is controlled by \(\theta_{k}\), to the observed curves \(\mathbf{Z}_{i}\):
\[
\mathbf{Z}_{b,\theta_{k},i} = \mathbf{Z}_{i} + \sqrt{\theta_{k}}\,\boldsymbol{\epsilon}_{b,i}, \qquad i = 1,\dots,n,
\]
where \(\boldsymbol{\epsilon}_{b,i}\) are generated as iid normal with mean \(\mathbf{0}\) and variance \(\hat{\sigma}_{\epsilon}^{2}\mathbf{I}\).
\item Obtain the coefficient function estimate \(\hat{\boldsymbol{\beta}}_{b,k}\) by
\[
\hat{\boldsymbol{\beta}}_{b,k} = F(\{\mathbf{Z}_{b,\theta_{k},i}, Y_{i}\}_{i=1,\dots,n}).
\]
\end{enumerate}
\item Take the average of \(B\) simulations to obtain the coefficient function estimate \(\hat{\boldsymbol{\beta}}_{k}\), for each \(\theta_{k} \in \{\theta_{1}, \dots, \theta_{K}\}\).
\item Model \(\hat{\boldsymbol{\beta}}_{k}\) in terms of \(\theta_{k}\). Based on the model, extrapolate back to \(\theta_{0} = -1\), which corresponds to the case of zero-error.
\end{enumerate}

\subsection*{Proof of Theorem 1}

In the simulation step, for a given \(\theta\), the estimated vector of B-spline coefficients \(\hat{\gamma}_{\theta}\) based on criterion (11) is obtained as
\[
\hat{\gamma}_{\theta} = (\mathbf{B}^{T}\mathbf{Z}_{\theta}^{T}\mathbf{Z}_{\theta}\mathbf{B} + \rho_{n}\mathbf{R})^{-1}\mathbf{B}^{T}\mathbf{Z}_{\theta}^{T}\mathbf{Y}.
\]
Then,
\[
\begin{aligned}
n^{1/2}(\hat{\gamma}_{\theta} - \gamma_{\theta})
&= n^{1/2}(\mathbf{B}^{T}\mathbf{Z}_{\theta}^{T}\mathbf{Z}_{\theta}\mathbf{B} + \rho_{n}\mathbf{R})^{-1}[\mathbf{B}^{T}\mathbf{Z}_{\theta}^{T}\mathbf{Y} - (\mathbf{B}^{T}\mathbf{Z}_{\theta}^{T}\mathbf{Z}_{\theta}\mathbf{B} + \rho_{n}\mathbf{R})\gamma_{\theta}] \\
&= n^{1/2}(\mathbf{B}^{T}\mathbf{Z}_{\theta}^{T}\mathbf{Z}_{\theta}\mathbf{B} + \rho_{n}\mathbf{R})^{-1}[\mathbf{B}^{T}\mathbf{Z}_{\theta}^{T}(\mathbf{Y} - \mathbf{Z}_{\theta}\mathbf{B}\gamma_{\theta}) - \rho_{n}\mathbf{R}\gamma_{\theta}] \\
&= \left( n^{-1}\sum_{i=1}^{n} (\mathbf{Z}_{\theta}\mathbf{B})_{i}^{T}(\mathbf{Z}_{\theta}\mathbf{B})_{i} + n^{-1}\rho_{n}\mathbf{R} \right)^{-1} \\
&\quad \times \left[ n^{-1/2}\sum_{i=1}^{n} (\mathbf{Z}_{\theta}\mathbf{B})_{i}^{T}(Y_{i} - (\mathbf{Z}_{\theta}\mathbf{B})_{i}\gamma_{\theta}) - n^{-1/2}\rho_{n}\mathbf{R}\gamma_{\theta} \right],
\end{aligned}
\]
where \((\mathbf{Z}_{\theta}\mathbf{B})_{i}\) is the \(i\)-th row of the matrix \(\mathbf{Z}_{\theta}\mathbf{B}\), with \(E((\mathbf{Z}_{\theta}\mathbf{B})_{ij}) = 0\) and \(E((\mathbf{Z}_{\theta}\mathbf{B})_{ij})^{2} < \infty\) (the latter is by Cauchy-Schwarz inequality), for \(i = 1,\dots,n\) and \(j = 1,\dots,p\). Since \(\rho_{n} = o_{p}(n^{1/2})\),
\[
n^{-1}\sum_{i=1}^{n} (\mathbf{Z}_{\theta}\mathbf{B})_{i}^{T}(\mathbf{Z}_{\theta}\mathbf{B})_{i} + n^{-1}\rho_{n}\mathbf{R} \xrightarrow{p} E[(\mathbf{Z}_{\theta}\mathbf{B})_{i}^{T}(\mathbf{Z}_{\theta}\mathbf{B})_{i}],
\]
and
\[
n^{-1/2}\rho_{n}\mathbf{R}\gamma_{\theta} \xrightarrow{p} 0,
\]
and \(n^{-1/2}\sum_{i=1}^{n} (\mathbf{Z}_{\theta}\mathbf{B})_{i}^{T}(Y_{i} - (\mathbf{Z}_{\theta}\mathbf{B})_{i}\gamma_{\theta})\) converges in distribution to a normal distribution. Then
\[
n^{1/2}(\hat{\gamma}_{\theta} - \gamma_{\theta}) = n^{-1/2}\sum_{i=1}^{n} \boldsymbol{\Delta}_{\theta}^{-1}(\mathbf{Z}_{\theta}\mathbf{B})_{i}^{T}(Y_{i} - (\mathbf{Z}_{\theta}\mathbf{B})_{i}\gamma_{\theta}) + o_{p}(1),
\]
where
\[
\boldsymbol{\Delta}_{\theta}^{-1} = E[(\mathbf{Z}_{\theta}\mathbf{B})_{i}^{T}(\mathbf{Z}_{\theta}\mathbf{B})_{i}] = \mathbf{B}^{T}(E(\mathbf{Z}_{i}^{T}\mathbf{Z}_{i}) + \theta\sigma_{\epsilon}^{2}\mathbf{I}_{D})\mathbf{B}.
\]

The above simulation step is repeated \(B\) times. Denote the generated functional predictors with additional measurement error at the \(b\)-th simulation by a \(n\times D\) matrix \(\mathbf{Z}_{b,\theta}\), and the corresponding vector of B-spline coefficient estimates by \(\hat{\gamma}_{b,\theta}\). Then
\[
n^{1/2}(\hat{\gamma}_{b,\theta} - \gamma_{\theta}) = n^{-1/2}\sum_{i=1}^{n} \boldsymbol{\Delta}_{\theta}^{-1}(\mathbf{Z}_{b,\theta}\mathbf{B})_{i}^{T}(Y_{i} - (\mathbf{Z}_{b,\theta}\mathbf{B})_{i}\gamma_{\theta}) + o_{p}(1), \qquad b = 1,\dots,B.
\]

For a given \(\theta\), the vector of B-spline coefficient estimates \(\hat{\gamma}_{S,\theta}\) is obtained by taking the average across simulations, \(\hat{\gamma}_{S,\theta} = \sum_{b=1}^{B}\hat{\gamma}_{b,\theta} / B\). Then
\[
n^{1/2}(\hat{\gamma}_{S,\theta} - \gamma_{\theta}) = n^{-1/2}\sum_{i=1}^{n} \sum_{b=1}^{B} \boldsymbol{\Delta}_{\theta}^{-1}(\mathbf{Z}_{b,\theta}\mathbf{B})_{i}^{T}(Y_{i} - (\mathbf{Z}_{b,\theta}\mathbf{B})_{i}\gamma_{\theta}) / B + o_{p}(1).
\]

The step above is performed for a sequence of \(\theta_{1},\dots,\theta_{K}\). Denote the true coefficient vector by \(\gamma_{\Theta} = \operatorname{vec}(\gamma_{\theta_{k}}, k = 1,\dots,K)\). Then, by the multivariate CLT,
\[
n^{1/2}(\hat{\boldsymbol{\gamma}}_{S,\Theta} - \gamma_{\Theta}) \xrightarrow{d} N(\mathbf{0}, \boldsymbol{\Lambda}^{-1}\boldsymbol{\Xi}\boldsymbol{\Lambda}^{-T}),
\]
where
\[
\boldsymbol{\Lambda} = \oplus_{k=1,\dots,K} \boldsymbol{\Delta}_{\theta_{k}},
\]
and
\[
\boldsymbol{\xi} = \operatorname{vec}\left( \sum_{b=1}^{B}(\mathbf{Z}_{b,\theta_{k}}\mathbf{B})_{i}^{T}(Y_{i} - (\mathbf{Z}_{b,\theta_{k}}\mathbf{B})_{i}\gamma_{\theta_{k}}) / B, \ k = 1,\dots,K \right),
\]
and \(\boldsymbol{\Xi} = \operatorname{Cov}(\boldsymbol{\xi})\) (here, \(\oplus\) is direct sum).

In the extrapolation step, based on criterion (12), the estimate \(\operatorname{vec}(\hat{\mathbf{A}}^{T})\) is given by
\[
\begin{aligned}
\operatorname{vec}(\hat{\mathbf{A}}^{T})
&= [(\boldsymbol{\Theta}\otimes \mathbf{B})^{T}(\boldsymbol{\Theta}\otimes \mathbf{B})]^{-1}(\boldsymbol{\Theta}\otimes \mathbf{B})^{T}(\mathbf{I}_{K}\otimes \mathbf{B})\hat{\boldsymbol{\gamma}}_{S,\Theta} \\
&= [(\boldsymbol{\Theta}\otimes \mathbf{B})^{T}(\boldsymbol{\Theta}\otimes \mathbf{B})]^{-1}(\boldsymbol{\Theta}\otimes \mathbf{B})^{T}(\mathbf{I}_{K}\otimes \mathbf{B})(\hat{\boldsymbol{\gamma}}_{S,\Theta} - \gamma_{\Theta}) + \operatorname{vec}(\mathbf{A}^{T}).
\end{aligned}
\]
Thus,
\[
\begin{aligned}
n^{1/2}\left( \operatorname{vec}(\hat{\mathbf{A}}^{T}) - \operatorname{vec}(\mathbf{A}^{T}) \right)
&= [(\boldsymbol{\Theta}\otimes \mathbf{B})^{T}(\boldsymbol{\Theta}\otimes \mathbf{B})]^{-1}(\boldsymbol{\Theta}\otimes \mathbf{B})^{T}(\mathbf{I}_{K}\otimes \mathbf{B})\left[ n^{1/2}(\hat{\boldsymbol{\gamma}}_{S,\Theta} - \gamma_{\Theta}) \right] \\
&\xrightarrow{d} N(\mathbf{0}, \mathbf{W}),
\end{aligned}
\]
in which
\[
\mathbf{V} = [(\boldsymbol{\Theta}\otimes \mathbf{B})^{T}(\boldsymbol{\Theta}\otimes \mathbf{B})]^{-1}(\boldsymbol{\Theta}\otimes \mathbf{B})^{T}(\mathbf{I}_{K}\otimes \mathbf{B}),
\]
\[
\mathbf{W} = \mathbf{V}\boldsymbol{\Lambda}^{-1}\boldsymbol{\Xi}\boldsymbol{\Lambda}^{-T}\mathbf{V}^{T}.
\]
Then, the SIMEX estimator \(\hat{\beta}_{\mathrm{SIMEX}} = \mathbf{B}\hat{\mathbf{A}}^{T}\boldsymbol{\Theta}_{0}^{T} = (\boldsymbol{\Theta}_{0}\otimes \mathbf{B})\operatorname{vec}(\hat{\mathbf{A}}^{T})\) has asymptotic variance \(\boldsymbol{\Sigma} = (\boldsymbol{\Theta}_{0}\otimes \mathbf{B})\mathbf{W}(\boldsymbol{\Theta}_{0}\otimes \mathbf{B})^{T}\).

\subsection*{Plots for Section~\ref{simulation}}

\begin{figure}[htbp]
\centering
\includegraphics[width=\linewidth]{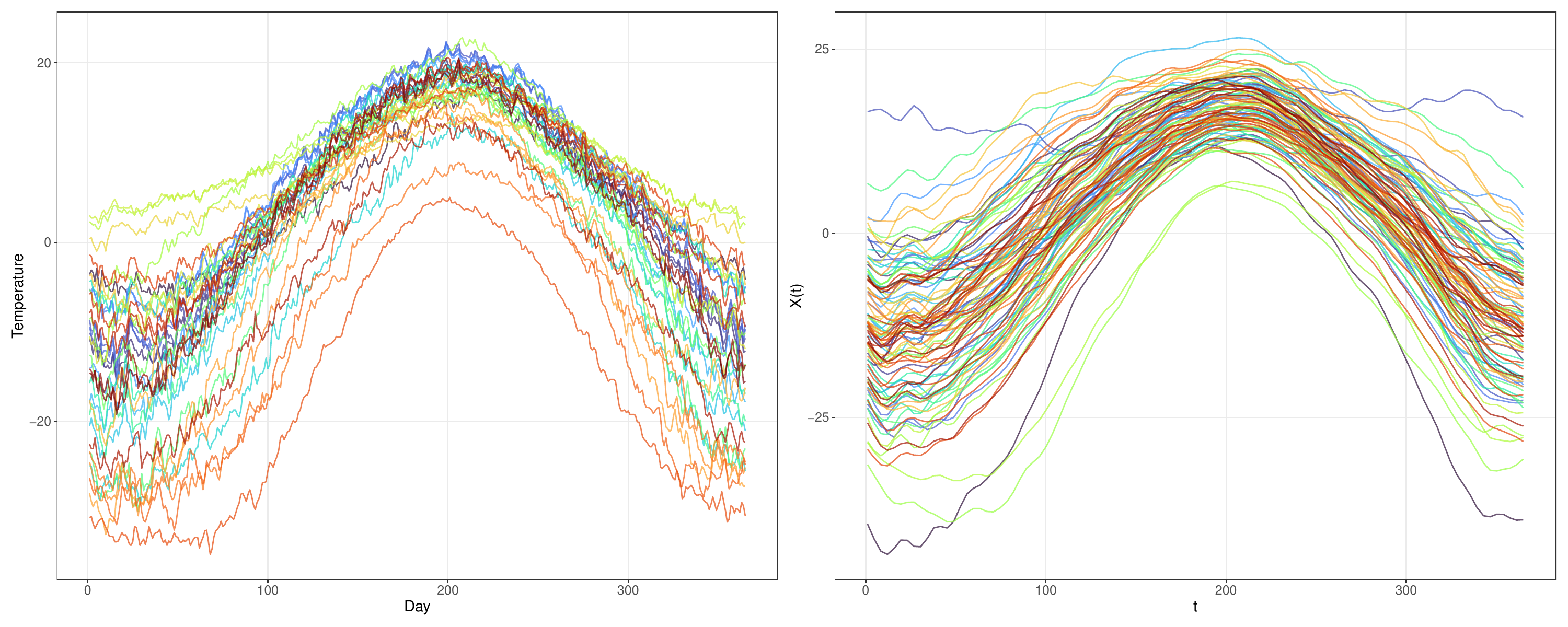}
\caption{Left: the plot of the original average daily temperature data over a year in the Canadian weather dataset. Curves observed at different locations are indicated by different colors. Right: the plot of simulated curves $X_{i}(t_{d})$ vs.~$t_{d}$, for $i=1,\dots,100$, from one of the $N=200$ samples.}
\label{fig:supp17}
\end{figure}

\begin{figure}[htbp]
\centering
\includegraphics[width=\linewidth]{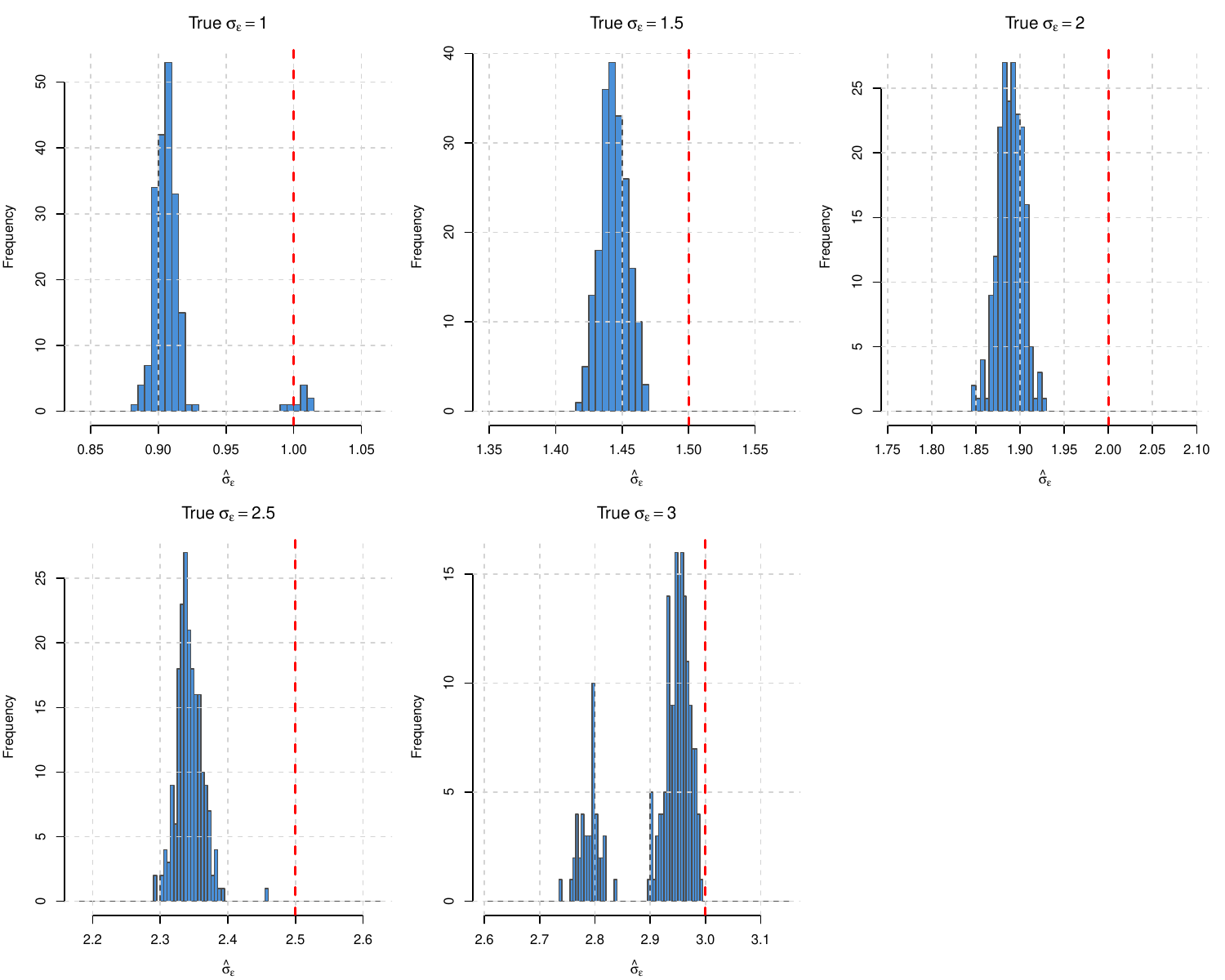}
\caption{Histograms of the estimated standard deviations of measurement error when the true standard deviation $\sigma_{\epsilon}$ is 1.0, 1.5, 2.0, 2.5, and 3.0 (from top left to bottom right). The dashed line indicates the true value.}
\label{fig:supp18}
\end{figure}

Figure~\ref{fig:supp17} shows the average daily temperature data at different locations over time based on the original Canadian weather dataset and an example of the simulated curves \(\mathbf{X}_{i}\). Figure~\ref{fig:supp18} gives histograms of the estimated standard deviations of measurement error.

\begin{thebibliography}{99}

\bibitem[Cardot et al.(2007)]{Cardot2007}
Cardot, H., Crambes, C., Kneip, A., and Sarda, P. (2007).
Smoothing splines estimators in functional linear regression with errors-in-variables.
\emph{Computational Statistics and Data Analysis} 51:4832--4848.

\bibitem[Cardot et al.(1999)]{Cardot1999}
Cardot, H., Ferraty, F., and Sarda, P. (1999).
Functional linear model.
\emph{Statistics and Probability Letters} 45:11--22.

\bibitem[Cardot et al.(2003)]{Cardot2003}
Cardot, H., Ferraty, F., and Sarda, P. (2003).
Spline estimators for the functional linear model.
\emph{Statistica Sinica} 13:571--591.

\bibitem[Carroll et al.(1996)]{Carroll1996}
Carroll, R. J., K{\"u}chenhoff, H., Lombard, F., and Stefanski, L. A. (1996).
Asymptotics for the SIMEX estimator in nonlinear measurement error models.
\emph{Journal of the American Statistical Association} 91:242--250.

\bibitem[Carroll et al.(2006)]{Carroll2006}
Carroll, R. J., Ruppert, D., Stefanski, L. A., and Crainiceanu, C. M. (2006).
\emph{Measurement Error in Nonlinear Models: A Modern Perspective}, Second Edition.
CRC Press, Boca Raton, FL.

\bibitem[Chen and Qiu(2023)]{Chen2023}
Chen, L.-P., and Qiu, B. (2023).
Analysis of length-biased and partly interval-censored survival data with mismeasured covariates.
\emph{Biometrics} 79(4):3600--3612.

\bibitem[Chen et al.(2024)]{Chen2024quantile}
Chen, X., Ji, H., Luan, Y., Zoh, R. S., Xue, L., Jadhav, S., and Tekwe, C. D. (2024).
Adjusting for bias due to measurement error in functional quantile regression models with error-prone functional and scalar covariates.
\emph{Biostatistics \& Epidemiology} 8(1):e2405439.

\bibitem[Cook and Stefanski(1994)]{CookStefanski1994}
Cook, J. R. and Stefanski, L. A. (1994).
Simulation-extrapolation estimation in parametric measurement error models.
\emph{Journal of the American Statistical Association} 89:1314--1328.

\bibitem[Crainiceanu et al.(2014)]{Crainiceanu2014}
Crainiceanu, C., Reiss, P., Goldsmith, J., Huang, L., Huo, L., Scheipl, F., Swihart, B., Greven, S.,
Harezlak, J., Kundu, M. G., Zhao, Y., McLean, M., and Xiao, L. (2014).
refund: Regression with Functional Data, R package version 0.1-10.

\bibitem[Craven and Wahba(1979)]{CravenWahba1979}
Craven, P. and Wahba, G. (1979).
Smoothing noisy data with spline functions: Estimating the correct degree of smoothing by the method of generalized cross-validation.
\emph{Numerische Mathematik} 31:377--403.

\bibitem[Davison and Hinkley(1997)]{DavisonHinkley1997}
Davison, A. C. and Hinkley, D. V. (1997).
\emph{Bootstrap Methods and their Application}.
Cambridge University Press, Cambridge.

\bibitem[Dempster et al.(1977)]{Dempster1977}
Dempster, A. P., Laird, N. M., and Rubin, D. B. (1977).
Maximum likelihood from incomplete data via the EM algorithm.
\emph{Journal of the Royal Statistical Society, Series B} 39:1--22.

\bibitem[Di et al.(2009)]{Di2009}
Di, C. Z., Crainiceanu, C., Caffo, B., and Punjabi, N. (2009).
Multilevel functional principal component analysis.
\emph{The Annals of Applied Statistics} 3:458--488.

\bibitem[Diggle and Hutchinson(1989)]{DiggleHutchinson1989}
Diggle, P. J. and Hutchinson, M. F. (1989).
On spline smoothing with autocorrelated errors.
\emph{Australian Journal of Statistics} 31:166--182.

\bibitem[Efron and Tibshirani(1994)]{EfronTibshirani1994}
Efron, B. and Tibshirani, R. J. (1994).
\emph{An Introduction to the Bootstrap}.
CRC Press, Boca Raton, FL.

\bibitem[Fan and Gijbels(1996)]{FanGijbels1996}
Fan, J. and Gijbels, I. (1996).
\emph{Local Polynomial Modelling and Its Applications}.
Chapman \& Hall, London.

\bibitem[Fuller(1987)]{Fuller1987}
Fuller, W. A. (1987).
\emph{Measurement Error Models}.
John Wiley \& Sons, New York.

\bibitem[Goldberg-Zimring et al.(2005)]{GoldbergZimring2005}
Goldberg-Zimring, D., Mewes, A. U. J., Maddah, M., and Warfield, S. K. (2005).
Diffusion tensor magnetic resonance imaging in multiple sclerosis.
\emph{Journal of Neuroimaging} 15:68S--81S.

\bibitem[Goldsmith et al.(2011a)]{Goldsmith2011a}
Goldsmith, J., Bobb, J., Crainiceanu, C. M., Caffo, B., and Reich, D. (2011a).
Penalized functional regression.
\emph{Journal of Computational and Graphical Statistics} 20:830--851.

\bibitem[Goldsmith et al.(2014)]{Goldsmith2014}
Goldsmith, J., Huang, L., and Crainiceanu, C. M. (2014).
Smooth scalar-on-image regression via spatial Bayesian variable selection.
\emph{Journal of Computational and Graphical Statistics} 23:46--64.

\bibitem[Goldsmith et al.(2011b)]{Goldsmith2011b}
Goldsmith, J., Wand, M. P., and Crainiceanu, C. M. (2011b).
Functional regression via variational Bayes.
\emph{Electronic Journal of Statistics} 5:572--602.

\bibitem[Gonzalez-Manteiga and Martinez-Calvo(2011)]{GonzalezManteiga2011}
Gonzalez-Manteiga, W. and Martinez-Calvo, A. (2011).
Bootstrap in functional linear regression.
\emph{Journal of Statistical Planning and Inference} 141:453--461.

\bibitem[Green and Silverman(1994)]{GreenSilverman1994}
Green, P. J. and Silverman, B. W. (1994).
\emph{Nonparametric Regression and Generalized Linear Models: A Roughness Penalty Approach}.
CRC Press, Boca Raton, FL.

\bibitem[Greven et al.(2010)]{Greven2010}
Greven, S., Crainiceanu, C., Caffo, B., and Reich, D. (2010).
Longitudinal functional principal component analysis.
\emph{Electronic Journal of Statistics} 4:1022--1054.

\bibitem[James(2002)]{James2002}
James, G. M. (2002).
Generalized linear models with functional predictors.
\emph{Journal of the Royal Statistical Society, Series B} 64:411--432.

\bibitem[James et al.(2009)]{James2009}
James, G. M., Wang, J., and Zhu, J. (2009).
Functional linear regression that's interpretable.
\emph{Annals of Statistics} 37:2083--2108.

\bibitem[Kohn et al.(1992)]{Kohn1992}
Kohn, R., Ansley, C. F., and Wong, C. (1992).
Nonparametric spline regression with autoregressive moving average errors.
\emph{Biometrika} 79:335--346.

\bibitem[Krivobokova and Kauermann(2007)]{Krivobokova2007}
Krivobokova, T. and Kauermann, G. (2007).
A note on penalized spline smoothing with correlated errors.
\emph{Journal of the American Statistical Association} 102:1328--1337.

\bibitem[Luan et al.(2023)]{Luan2023a}
Luan, Y., Zoh, R. S., Jadhav, S., Xue, L., and Tekwe, C. D. (2023).
Generalized functional linear regression models with a mixture of complex function-valued and scalar-valued covariates prone to measurement error.
\emph{arXiv preprint} arXiv:2304.02651.


\bibitem[Marx and Eilers(1999)]{MarxEilers1999}
Marx, B. and Eilers, P. (1999).
Generalized linear regression on sampled signals and curves: A P-spline approach.
\emph{Technometrics} 41:1--13.

\bibitem[McKeague and Sen(2010)]{McKeagueSen2010}
McKeague, I. W. and Sen, B. (2010).
Fractals with point impact in functional linear regression.
\emph{Annals of Statistics} 38:2559--2586.

\bibitem[Nadaraya(1964)]{Nadaraya1964}
Nadaraya, E. A. (1964).
On estimating regression.
\emph{Theory of Probability and its Applications} 9:141--142.

\bibitem[Opsomer et al.(2001)]{Opsomer2001}
Opsomer, J., Wang, Y., and Yang, Y. (2001).
Nonparametric regression with correlated errors.
\emph{Statistical Science} 16:134--153.

\bibitem[Ozturk et al.(2010)]{Ozturk2010}
Ozturk, A., Smith, S. A., Gordon-Lipkin, E. M., Harrison, D. M., Shisee, N., Pham, D. L., Caffo, B. S., Calabresi, P. A., and Reich, D. S. (2010).
MRI of the corpus callosum in multiple sclerosis: Association with disability.
\emph{Multiple Sclerosis} 16:166--177.

\bibitem[Ramsay et al.(2009)]{Ramsay2009}
Ramsay, J. O., Hooker, G., and Graves, S. (2009).
\emph{Functional Data Analysis with R and MATLAB}.
Springer, New York.

\bibitem[Ramsay and Silverman(2005)]{RamsaySilverman2005}
Ramsay, J. O. and Silverman, B. W. (2005).
\emph{Functional Data Analysis}, Second Edition.
Springer, New York.

\bibitem[Ramsay et al.(2014)]{Ramsay2014}
Ramsay, J. O., Wickham, H., Graves, S., and Hooker, G. (2014).
fda: Functional Data Analysis, R package version 2.4.3.

\bibitem[Reiss and Ogden(2010)]{ReissOgden2010}
Reiss, P. and Ogden, R. T. (2010).
Functional generalized linear models with images as predictors.
\emph{Biometrics} 66:61--69.

\bibitem[Reiss et al.(2016)]{Reiss2016}
Reiss, P. T., Goldsmith, J., Shang, H. L., and Ogden, R. T. (2016).
Methods for scalar-on-function regression.
\emph{International Statistical Review}.

\bibitem[Reiss et al.(2010)]{ReissHuang2010}
Reiss, P. T., Huang, L., and Mennes, M. (2010).
Fast function-on-scalar regression with penalized basis expansions.
\emph{International Journal of Biostatistics} 6:Article 28.

\bibitem[Reiss and Ogden(2007)]{ReissOgden2007}
Reiss, P. T. and Ogden, R. T. (2007).
Functional principal component regression and functional partial least squares.
\emph{Journal of the American Statistical Association} 102:984--996.

\bibitem[Ruppert et al.(2003)]{Ruppert2003}
Ruppert, D., Wand, M., and Carroll, R. (2003).
\emph{Semiparametric Regression}.
Cambridge University Press, Cambridge, U.K.

\bibitem[Scheipl and Greven(2012)]{ScheiplGreven2012}
Scheipl, F. and Greven, S. (2012).
Identifiability in penalized function-on-function regression models.
Technical Report Number 125, LMU.

\bibitem[Sevilimedu and Yu(2022)]{Sevilimedu2022}
Sevilimedu, V., and Yu, L. (2022).
Simulation extrapolation method for measurement error: A review.
\emph{Statistical Methods in Medical Research} 31(8):1617--1636.

\bibitem[Staniswalis and Lee(1998)]{StaniswalisLee1998}
Staniswalis, J. G. and Lee, J. J. (1998).
Nonparametric regression analysis of longitudinal data.
\emph{Journal of the American Statistical Association} 93:1403--1418.

\bibitem[Van~Huffel and Vandewalle(1991)]{VanHuffel1991}
Van~Huffel, S. and Vandewalle, J. (1991).
\emph{The Total Least Squares Problem: Computational Aspects and Analysis}.
SIAM, Philadelphia.

\bibitem[Wahba(1983)]{Wahba1983}
Wahba, G. (1983).
Bayesian confidence intervals for the cross-validated smoothing spline.
\emph{Journal of the Royal Statistical Society, Series B} 45:133--150.

\bibitem[Wahba and Wang(1995)]{WahbaWang1995}
Wahba, G. and Wang, Y. (1995).
Behavior near zero of the distribution of GCV smoothing parameter estimates.
\emph{Statistics \& Probability Letters} 24:105--111.

\bibitem[Wang(1998)]{Wang1998}
Wang, Y. (1998).
Smoothing spline models with correlated random errors.
\emph{Journal of the American Statistical Association} 93:341--348.

\bibitem[Watson(1964)]{Watson1964}
Watson, G. S. (1964).
Smooth regression analysis.
\emph{Sankhya: the Indian Journal of Statistics (Series A)} 26:359--372.

\bibitem[Wood(2006a)]{Wood2006a}
Wood, S. N. (2006a).
\emph{Generalized Additive Models: An Introduction with R}.
CRC Press, Boca Raton, FL.

\bibitem[Wood(2006b)]{Wood2006b}
Wood, S. N. (2006b).
Low rank scale invariant tensor product smooths for generalized additive mixed models.
\emph{Biometrics} 62:1025--1036.

\bibitem[Wood(2011)]{Wood2011}
Wood, S. N. (2011).
Fast stable restricted maximum likelihood and marginal likelihood estimation of semiparametric generalized linear models.
\emph{Journal of the Royal Statistical Society, Series B} 73:3--36.

\bibitem[Yao et al.(2003)]{Yao2003}
Yao, F., M{\"u}ller, H.-G., Clifford, A. J., Dueker, S. R., Follett, J., Lin, Y., Buchholz, B. A., and Vogel, J. S. (2003).
Shrinkage estimation for functional principal component scores with application to the population kinetics of plasma folate.
\emph{Biometrics} 59:676--685.

\bibitem[Yao et al.(2005)]{Yao2005}
Yao, F., M{\"u}ller, H.-G., and Wang, J.-L. (2005).
Functional data analysis for sparse longitudinal data.
\emph{Journal of the American Statistical Association} 100:577--591.

\bibitem[Zoh et al.(2024)]{Zoh2024}
Zoh, R. S., Luan, Y., Xue, L., Allison, D. B., and Tekwe, C. D. (2024).
A Bayesian semi-parametric scalar-on-function regression with measurement error using instrumental variables.
\emph{Statistics in Medicine} 43(21):4043--4054.

\end{thebibliography}
\end{document}